\documentclass[showkeys,prd,nofootinbib,floatfix,preprintnumbers]{revtex4-1}

\usepackage[utf8]{inputenc}
\usepackage{multirow}
\usepackage{amsfonts,amsmath,amssymb} 
\usepackage{graphicx,graphics,color}
\usepackage{gensymb}
\usepackage{longtable}
\usepackage{bbding}
\usepackage{subfigure}
\usepackage[T1]{fontenc}
\usepackage{hyperref}
\usepackage[normalem]{ulem}
\hypersetup{
    colorlinks=true,
    linkcolor=blue}

\begin{document}




\title{Non-linear causal bulk viscosity in Unified Dark Matter Cosmologies}

\author{Guillermo Palma}
\email{guillermo.palma@usach.cl}
\affiliation{Departamento de F\'isica, Universidad de Santiago de Chile,\\Avenida V\'ictor Jara 3493, Estaci\'on Central, 9170124, Santiago, Chile}

\author{Gabriel G\'omez}
\email{luis.gomezd@umayor.cl}
\affiliation{Centro Multidisciplinario de F\'isica, Vicerrector\'ia de Investigaci\'on, Universidad Mayor, \\ Camino La Pir\'amide 5750,  Huechuraba, 8580745, Santiago, Chile}

\date{\today}

\begin{abstract}
We propose a bulk viscous unified dark matter scenario based on a non-linear extension of the full causal Israel–Stewart theory. This framework allows the viscous fluid to remain far from equilibrium—an essential feature for a physically consistent description of viscosity-driven accelerated expansion. We adopt the standard parametrization for the bulk viscosity, $\xi=\xi_{0}\ \rho_{m}^{s}$, treating $s$ as a free parameter (in contrast to most previous works), and study the model in a spatially flat Friedmann–Robertson–Walker background. By reformulating the cosmological equations as an autonomous dynamical system, we obtain both asymptotic analytical solutions and a numerical characterization of the phase space.  At early times, the viscous component can mimic a stiff fluid, while at intermediate epochs it behaves like dark matter. With a suitable choice of dynamical variables, the system admits three distinct classes of late-time attractors. Two of them are separated by a basin-boundary saddle point: (i) a generic quintessence solution for $s = 1/2$, which encompasses a de Sitter–like behavior when $\xi_{0}$ satisfies a specific relation involving the nonlinear parameters; (ii) a global exact de Sitter attractor for $s < 1/2$; and (iii) a phantom-like solution that emerges for $s \geq 1/2$. In contrast to the generic $s\neq 1/2$ case, the $s=1/2$ scenario exhibits a qualitatively different stability structure, allowing de Sitter and phantom attractors to coexist. All solutions respect entropy production, and cosmic acceleration emerges independently of $\xi_{0}$, relaxing the strong bounds, $\xi_{0}\sim\mathcal{O}(1)$, required in Eckart-based viscous models. 
\end{abstract}

\pacs{Valid PACS appear here}
\maketitle

\section{Introduction}

Our current understanding of the Universe is reflected in the standard cosmological model, known as $\Lambda$CDM. This framework includes a cosmological constant ($\Lambda$) to account for the observed accelerated expansion of the Universe, and a cold dark matter (CDM) component responsible for shaping the large-scale structures \cite{Peebles:1994xt,Planck:2018vyg}. Despite its success, growing tensions with observational data have raised legitimate concerns about the completeness of the $\Lambda$CDM model \cite{Abdalla:2022yfr}.

Recently, combined analyses of the Dark Energy Spectroscopic Instrument (DESI) Data Release 2 of the Baryon Acoustic Oscillations (BAO) \cite{DESI:2025fii}, the cosmic microwave background (CMB) from the Planck satellite \cite{Planck:2018vyg}, and Type Ia supernovae (SNIa) \cite{Scolnic:2021amr,Brout:2022vxf,DES:2024jxu} have provided strong evidence suggesting a time-evolving dark energy component at low redshift \cite{DESI:2025zgx}. This adds to existing discrepancies within the model, such as the well-known Hubble tension \cite{Riess:2021jrx,Knox:2019rjx} and the $\sigma_8$ tension \cite{Macaulay:2013swa,Battye:2014qga,Alam:2016hwk,Abbott:2017wau}. Furthermore, a longstanding issue—commonly referred to as the small-scale crisis—highlights significant disagreements between $N$-body simulations and observations at kiloparsec scales, particularly in dwarf galaxy structures (see e.g. \cite{Bullock:2017xww}).

Reconciling theoretical predictions with observational data is a central challenge in modern cosmology. Among the various proposed solutions to the small-scale crisis is the idea of modifying the nature of dark matter itself. One promising approach involves introducing self-interacting dark matter \cite{Spergel:1999mh}, which can address small-scale anomalies while preserving the successful large-scale predictions of the CDM paradigm. These self-interactions may give rise to a macroscopic bulk viscous pressure, interpreted as a manifestation of dissipative process \cite{Atreya:2017pny,Mishra:2020onx,Gagnon:2011id}.

Such a non-perfect fluid could impact not only (sub)galactic structure formation but also contribute directly to the current accelerated expansion of the Universe. In this context, abandoning the idea of a cosmological constant allows the bulk viscous pressure to take on the role of driving cosmic acceleration—effectively acting as a form of repulsive gravity. This idea gives rise to the so-called viscous unified dark matter models \cite{Padmanabhan:1987dg,Bulk1,Bulk2,Colistete:2007xi,Brevik:2017msy,blas2015large,Barbosa:2017ojt,foot2016solving,Bulk3,Bulk4,Bulk5,bulk6}, which belong to a broader class of unified dark sector scenarios aiming to describe both dark matter and dark energy as different manifestations of a single fluid.

From a microscopic perspective, several mechanisms have been proposed to explain the origin of bulk viscosity at cosmological scales. For instance \cite{Zimdahl:1996fj} shows that two fluid components with different cooling rates can generate an effective bulk viscosity for the combined system to reach thermal equilibrium. Thermal field theory approaches introduce self-interacting scalar fields to model dark energy with viscous effects \cite{Gagnon:2011id}, while the connection between particle creation and bulk viscosity—particularly relevant during inflation—is explored in \cite{Brevik:1996ca,Murphy:1973zz,1982PhLA...90..375H,Eshaghi:2015tqa,Bamba:2015sxa}. An alternative microscopic mechanism considers bulk viscosity arising from dark matter annihilation, contributing to late-time cosmic acceleration \cite{Wilson:2006gf, Mathews:2008hk}. Additionally, kinetic theory has been applied to model viscous effects in self-interacting dark matter scenarios \cite{Atreya:2017pny, Natwariya:2019fif}. In the context of neutralino CDM, collisional damping during kinetic decoupling leads to energy dissipation from dark matter to radiation \cite{Hofmann:2001bi}. These mechanisms illustrate the potential microscopic origins of viscous effects relevant for the Universe’s dynamical evolution.

At this point, it is relevant to mention classical formulations used to describe relativistic dissipative fluids. At first order, Eckart’s theory \cite{Eckart} (see also \cite{LandauandLifshitz}) provides a simplified framework within relativistic hydrodynamics to account for macroscopic dissipative effects. However, this theory suffers from serious drawbacks, such as unstable equilibrium states \cite{Hiscock:1985zz} and superluminal propagation of high-frequency modes \cite{Hiscock:1987zz}. These pathologies stem from the fact that the corresponding set of partial differential equations is not hyperbolic, preventing a well-posed Cauchy initial value problem. These issues were resolved in the Müller–Israel–Stewart (MIS) theory \cite{Muller:1967zza,Israel:1976tn,Israel:1976efz}, which introduces independent relaxation dynamics for the additional degrees of freedom (dissipative fields), effectively modifying the theory’s high-frequency behavior. The equations of motion for the new variables were originally postulated based on the second law of thermodynamics \cite{Israel:1976tn}, differing from more modern formulations of relativistic hydrodynamics \cite{Baier:2007ix,Denicol:2012cn}. The MIS theory is fully causal, stable, and admits a well-posed initial value formulation even when dynamically coupled to gravity \cite{Bemfica:2019cop}. As in Eckart’s theory, the original MIS formulation assumes only small deviations from equilibrium, leading to a transport equation that is linear in the bulk viscous pressure. 

A non-linear extension of bulk viscosity was proposed by Maartens et al. in Ref.~\cite{Maartens:1996dk}, who introduced non-linear effects phenomenologically through a characteristic time scale, $\tau_{*}$. In this extended framework, the near-equilibrium assumption is relaxed, enabling a more general and physically realistic treatment of dissipative processes. As a result, the viscous fluid is allowed to deviate significantly from equilibrium—a necessary condition for modeling viscosity-driven cosmic acceleration. A key feature of this non-linear extension is that it recovers the standard MIS theory in the linear regime ($\tau_{*} \to 0$).

Most existing bulk viscous unified models are based on Eckart’s theory, which, despite being pathological, remains a practical framework for modeling the dynamics of the cosmic dark fluid. See e.g. \cite{Bulk2,Colistete:2007xi,Palma:2024qrw,Yang:2019qza}. Other studies have instead relied on truncated or linearized versions of the IS theory to explore the cosmological properties of a unified cosmic fluid \cite{Piattella:2011bs,Mohan:2017poq,Cruz:2018psw,Cruz:2019uya} (see also \cite{Disconzi:2014oda} for a different approach to model bulk viscosity). These investigations range from analytical solutions in Friedmann–Robertson–Walker (FRW) backgrounds to linear stability analyses of dynamical fixed points. The main argument supporting such approximations is that the nonlinear complexity of the full theory prevents analytical expressions for the Hubble parameter and limits insight into the dynamical evolution of the Universe. However, as we demonstrate in this work, this limitation does not always apply. Moreover, truncated causal and non causal formulations may yield results that deviate significantly from those of the full causal theory, potentially leading to misleading conclusions. 

In this work, we aim to fill a critical gap by implementing the full nonlinear Müller–Israel–Stewart framework within the context of a bulk viscous unified dark matter model. In this regard, our approach departs significantly from Ref.~\cite{Aguilar-Perez:2022bzb}, where an explicit dark energy component was introduced to drive late-time accelerated expansion; or from  \cite{Acquaviva:2015hsa}, which relied on a hypothetical viscous radiation component lacking observational support. In these models, dissipation in the dark matter sector is treated only as a secondary effect. Accordingly, our objectives are twofold: (i) to 
extend no-lineal bulk viscosity in unified DM models and (ii) to investigate whether the inconsistencies found in previous approximate treatments are intrinsic to the theory or merely artifacts of the underlying simplifications. Together, these elements determine the model’s potential to explain the present accelerated cosmic expansion.

Our results reveal the presence of two late-time attractor solutions in the non-linear regime of the causal theory: one corresponding to an exact de Sitter–type solution and the other to a phantom regime. Both fixed points satisfy the non-negativity condition for entropy production. Each of these accelerated phases is preceded by a decelerated expansion era, during which structure formation can occur, implying that viscous effects must remain suppressed at intermediate times \cite{Anand:2017wsj}. Overall, this study represents a necessary step toward a more consistent and physically grounded description of cosmic dissipation in the dark sector within viscous unified dark matter scenarios.

Finally, let us mention that, to date, there is no widely accepted model that derives bulk viscosity from first principles at the microscopic level. As a result, many viscous unified dark matter models rely on phenomenological parametrizations for the bulk viscosity coefficient $\xi$, which plays a central role in dissipative cosmologies. The most common and natural choice—adopted in this work—is a power-law dependence on the matter energy density, $\xi \propto \rho_{m}^{s}$. The specific case $s=1/2$  has been extensively studied in the context of bulk viscous cosmology \cite{Big.Bang,rho1,Brevik,Analysing,valorxi1,valorxi2,primerarticulo,sym14091866,Cruz:2022zxe}, while the case $s=0$, corresponding to a constant dissipation, was considered in \cite{Velten:2011bg}, displaying problems at perturbative level. In contrast, we treat $s$ as a free parameter and carry out our study by considering separately the cases $s\neq 1/2$ and $s= 1/2$, since the latter is more conveniently handled through a different parametrization of the phase-space variables. 
 
This paper is structured as follows. In Section \ref{sec:II}, we summarize the essential features of the full causal non-linear model in a FRW background. In Section \ref{sec:III}, we reformulate the system as an autonomous dynamical system, which serves a dual purpose: enabling the derivation of analytical results and facilitating a numerical exploration of the phase space. The numerical analysis, along with a detailed discussion of the resulting trajectories and critical points, is presented in Section \ref{sec:IV}. Finally, in Section \ref{sec:V}, we summarize and discuss the main results, emphasizing their implications for cosmology and the potential of bulk viscous unified dark matter to explain the current phase of cosmic acceleration.

\section{Full Causal Viscous Model}\label{sec:II}
In this section, we will briefly present and discuss the fundamental equations describing the dynamics of the model including the modification of the full causal Israel-Stewart dissipative theory \cite{Israel:1976tn} within the Dark Matter (DM) sector in unified models. This nonlinear modification of the causal linear thermodynamics IS theory includes a generalized dependence of the bulk viscous pressure on the "thermodynamic force", which ensures both the positivity of the entropy and the non-negativity of entropy production, consistent with the second law of thermodynamics. It was first proposed in \cite{Maartens:1996dk}, and applied to an inflationary scenario.

This generalized relativistic viscosity approach represents a substantial improvement over the standard IS theory. Notably, it remains valid in situations where microscopic hydrodynamical processes, such as nonzero heat fluxes, would render the original IS formulation inapplicable \cite{PhysRevE.50.4233}.

Building on the findings of the authors who applied Eckart’s relativistic viscous theory to dissipative DM unified models \cite{Palma:2024qrw}, we will demonstrate that the nonlinear extension of IS full causal viscous theory leads to a dynamical equation for DM bulk viscosity pressure, whose solution evolves towards an effective negative pressure values giving rise to the current cosmic accelerated expansion. As in this former study, we will include radiation in the setup to illustrate later that, unlike the previous non-causal version, demanding a complete cosmological evolution is notably always possible in the present scenario. Further, and for the sake of simplicity, we will assume a homogeneous and isotropic background described by a flat Friedmann–Robertson–Walker (FRW) metric. The universe components are represented by two fluids: one for radiation with energy density $\rho_{r}$ and pressure $P_{r}$, and another for dark matter with energy density $\rho_{m}$ and a pressure $P_{m}$. The effective pressure $P_{m}^{\rm eff}$ will evolve according to the dynamical equation of the extended IS dissipative theory. The Friedmann constraint and the Raychaudhuri equations are respectively given by: 

\begin{eqnarray}
    3H^{2}&=&8\pi G_{N} \left(\rho_{r}+\rho_{m}\right),  \label{constraint}\\ 
    3H^{2}+2\dot{H}&=&-8\pi G_{N} \left(P_{r}+P_{m}^{\rm eff}\right),
    \label{Raychaudhuri}
\end{eqnarray}
where $G_N$ denotes the Newton's constant. We will assume a barotropic relation for both fluids, for radiation $P_{r}=\rho_{r}/3$, and for the DM assume a generalized fluid behavior with an EoS $P_{m}= (\gamma -1 )\rho$, with particular focus on the case $\gamma=1$, which corresponds to an intrinsically pressureless DM fluid.

As a consequence of the dissipative processes within the DM sector the pressure get an extra contribution called bulk viscous pressure $\Pi$, leading to an effective pressure defined by
\begin{equation}
    P_{m}^{\rm eff}=P_{m}+\Pi= (\gamma -1 )\rho_{m} + \Pi.\label{eff_pressure}
\end{equation}
The corresponding nonlinear equation for the bulk viscous pressure generalizes the Israel-Stewart transport equation, ensuring the fulfilment of the second Thermodynamics law and the positiveness of the effective specific entropy. It is given by the following expression:

\begin{equation}
   \tau \; \dot{\Pi} + 3 \; \xi H + \Pi \left(1+ \frac{\tau_*}{\xi} \; \Pi \right)^{-1} + \frac{1}{2} \; \tau \; \Pi \left[3 \; H + \frac{\dot{\tau}}{\tau} -\frac{\dot{\xi}}{\xi} - \frac{\dot{T}}{T} \right] = 0.\label{IS_modified}
\end{equation}
Here $\tau$ and $\tau_*$ are, respectively, the relaxation times characteristic for linear and nonlinear effects. From the above expression follows that in the limit $\tau_* =0$, the above transport equation goes into the IS expression. The linear relaxation time has been expressed as a function of the bulk viscosity in \cite{Maartens:1996vi} in the form

\begin{equation}
    \tau = \frac{\xi}{v^2 \; \gamma \; \rho_m}, \label{linear_relax_time}
\end{equation}
where $v$ is the dissipative contribution to the sound speed $V$, i. e. $V^2 = v^2 + v_s^2$, while $v_s$ is the adiabatic one. Due to the causality, $V \leq 1 $, $v^2 $ must be bounded from above, that is, $v^2 \leq (2 - \gamma)$.

For the nonlinear relaxation time we will also assume the simplest choice proposed in \cite{Maartens:1996dk}:
\begin{equation}
    \tau_* = k^2 \;\tau  , \label{nonlinear_relax_time}
\end{equation}
where $k$ is a real constant. In addition, $T$ denotes the common equilibrium temperature of the fluids, which, due to their barotropic nature, must take the functional form
\begin{equation}
    T = T_0 \; \rho_m^{(1-1/\gamma)}. \label{eff_temperature}
\end{equation}
As elaborated below Eq.~(\ref{positive_entropy}), the nonlinear relaxation time allows for a physically consistent description of far-from-equilibrium viscous dynamics in fluids. Finally, $\xi$ denotes the bulk viscosity coefficient, which, in principle, should be derived from first principles using kinetic theory \cite{Zimdahl:1996fj}. However, in the absence of a microscopically motivated expression for $\xi$, we adopt a phenomenological Ansatz of the form $\xi \propto \rho_m^s$, a simple power-law behavior that is widely used in the literature. We deliberately exclude any explicit dependence of $\xi$ on the Hubble parameter $H$ or its derivatives (e.g., $\dot{H}$), as such assumptions would implicitly introduce couplings between the viscous DM fluid and other cosmic components (see, e.g., \cite{Gomez:2022qcu} for further discussion). Explicitly, we use the following parametrization:  

\begin{equation}
    \xi \equiv \frac{\xi_{0}/\sqrt{3}}{(8\pi G_{N})^{(3/2-2s)}} \rho_{m}^{s}. 
\label{power_law_viscosity}
\end{equation}
This parametrization conveniently includes the widely studied case of constant bulk viscosity for $s=0$. It is also worth noting that $\xi_{0}$ remains a dimensionless parameter for arbitrary values of the viscous exponent $s$. According to the nonlinear generalization of the causal IS thermodynamics \cite{Maartens:1996dk}, the positiveness of the entropy production rate is guaranteed by the condition:
\begin{equation}
    \dot{S} = - \frac{3 H \Pi }{nT} \geq 0, \quad \text{or equivalently,} \quad -\Pi \leq \frac{\xi}{\tau_{*}},
\label{entropy_production}
\end{equation}
where $T$ is the local equilibrium temperature, $S$ is the specific entropy and $n$ is the particle number density in the local reference equilibrium state. It is defined in terms of the particle number four-current, $n^{\alpha}$, as $n^{\alpha} = n u^{\alpha}$, with $u^{\alpha}$ being the fluid four-velocity, and it is constrained in the same way as the energy-momentum tensor $T_{\alpha \beta}$ to the conservation condition $n^{\alpha}_{\; ;\alpha} = 0$, leading rise to the equation
\begin{equation}
    \dot{n} + 3 H n = 0.
\label{number_conservation}
\end{equation}
In addition, the positiveness of the effective specific entropy, $S_{\rm eff}$, leads to the condition 
\begin{equation}
    S_{\rm eff}  \geq 0, \quad \text{or equivalently,} \quad  \mid \Pi \mid \leq \mid \Pi_{max} \mid = \sqrt{\frac{2nTS\xi}{\tau}}. 
\label{positive_entropy}
\end{equation}
We highlight here a crucial point: in both the original Eckart theory and its causal IS extension, deviations from equilibrium are required, by construction, to remain small.
In the IS framework, this near-equilibrium condition ensures the positivity of the effective entropy and is typically expressed as $|\Pi| \ll p_{m}$. While this is generally not problematic for high-density fluids (such as those in neutron stars), it becomes a critical limitation in cosmology: for a dark matter component with $p_{m} \approx 0$, the above condition is inevitably violated (see, for instance, Ref.~\cite{Maartens:1996dk}). 

In contrast, within the nonlinear extension of the IS theory, the positivity of the effective entropy leads to a finite upper bound on the viscous pressure given by Eq.~(11). Importantly, this bound does not depend on the ``bare'' pressure $p_{m}$, thereby permitting substantially larger departures from equilibrium and allowing for a physically consistent description of far-from-equilibrium viscous dynamics in cosmological settings.

To complete the description of the system’s dynamics, the energy–momentum conservation laws for the two fluid components are required. These can be expressed in the following compact form:
\begin{align}
    \dot{\rho}_{r} + 4 H \rho_{r} =0,\label{continuity_rho_rad} \\ 
 \dot{\rho}_{m} + 
  3 H (\gamma \rho_{m} + \Pi) =0.
  \label{continuity_rho_DM}
\end{align}
We are now prepared to study the above set of coupled differential equations for arbitrary values of the parameter 
$s$ appearing in Eq.~(\ref{power_law_viscosity}). In the next section, we recast these equations into an autonomous differential system by introducing suitable definitions of the phase-space variables as dimensionless quantities. Rather than seeking numerical solutions, which would be highly challenging due to the strong nonlinearities and singularities in the evolution equations (particularly for $s>1/2$), we employ the dynamical systems approach to identify the fixed points and analyze their stability properties. Our aim is to test whether the present fully causal setup, applied to unified viscous DM models, can account not only for the recent accelerated expansion but also for the known stages of cosmic evolution.


\section{Autonomous Evolution Equations and Dynamical System Analysis}
\label{sec:III}
Now we introduce dimensionless variables that span the phase space of the system. The inclusion of the dynamical evolution equation for bulk viscosity (see Eq.~(\ref{IS_modified})) leads to an increase in the phase space dimension, when compared to the non-causal Eckart's viscous theory. In principle one could introduce the usual variables  

\begin{align}
& X \equiv 8\pi G_{N} ~H^2 \; ; \; \Omega_{r} \equiv \frac{8\pi G_{N}}{3~H^2} ~\rho_{r} ~; \Omega_{m} \equiv \frac{8\pi G_{N}}{3~H^2} ~\rho_{m} ~; \;\; Z \equiv \frac{8\pi G_{N}}{3~H^2}  \; \Pi .
\label{old_Phase_Space_variables}
\end{align}
These variables are clearly redundant, as the constraint of Eq.~(\ref{constraint}) implies $\Omega_{r} + \Omega_{m} = 1$. More importantly, however, the corresponding differential equations become ill-defined at certain physically relevant fixed points.
Therefore, we adopt the phase space variables $(U, V, W )$ defined as the following dimensionless ratios
\begin{align}
& U \equiv \frac{(8\pi G_{N})^{(3/2-2s)}}{3^{(1-s)}} \; \frac{\rho_{m}^{(1-s)}}{H} ; \;\; V \equiv \frac{\Pi}{\rho_{m}}; \;\; W \equiv \frac{8\pi G_{N}}{3}  \; \frac{\Pi}{H^2} .
\label{Phase_Space_variables}
\end{align}
It is worthwhile pointing out that for the particular case of Eckart's viscous theory $\Pi = -3 H \xi$, the above variables become redundant as the product $ U \;V $ goes into a constant, reducing the phase space dimensions to two. Accordingly, $V$ and $W$ might take either positives or negatives values but simultaneously. Indeed, since the dimensionless matter density parameter $\Omega_{m} = 8\pi G_{N} \rho_{m}/3 H^{2}$ is by definition strictly positive, and can equivalently be written as $\Omega_{m} = W/V$, it follows directly that $V$ and $W$ must necessarily share the same sign. Notice that $U$ is related to the dark matter energy density in a non-linear way, leading to a non-compact phase space. $V$ effectively corresponds to the equation of state of the viscous fluid, while $W$ denotes the normalized viscous pressure, quantifying the level of viscosity during the expansion. 
After some cumbersome analytical manipulations the dynamical system can be cast as follows:
\begin{equation}
U^\prime =  U \; \left[ -3 (1-s) \left( \; \gamma + V \right) + 2 - (4-3\gamma) \frac{W}{2V} + \frac{3}{2} \; W  \right] ,\label{evol_U}
\end{equation}
\begin{equation}
V^\prime = -3 v^2 \gamma  \left[ \; 1 + \frac{3^{(1/2-s)}}{\xi_0} U V \; \left(1 + \frac{k^2}{v^2 \gamma} V \right)^{-1} \right] +  \frac{3}{2 \gamma} \; V^2 , \label{evol_V}
\end{equation}
\begin{equation}
W^\prime = -3 v^2 \gamma  \left[ \; \frac{1}{V} + \frac{3^{(1/2-s)}}{\xi_0} U  \; \left(1 + \frac{k^2}{v^2 \gamma} V \right)^{-1} \right] W +  \left[ \; (4 - 3 \gamma ) \left(1- \frac{W}{V} \right) + 3 W - \frac{3}{2 \gamma} (2 \gamma - 1 ) V \right] W \label{evol_W},
\end{equation}
where the prime denotes derivative with respect to the number of \textit{e-folds}, $N\equiv\ln a$. These expressions explicitly show a dependence on the bulk viscosity through the constant $\xi_0$, which can be interpreted as a nonlinear self-interaction of the DM fluid \cite{Zimdahl:1996fj}. This feature leads to non-trivial effects on the background dynamics and gives rise to a non-vanishing bulk viscosity. Notice that the submanifold $V=0$ is not allowed as it leads to a singularity in the dynamical system. This, in principle, prevents the possibility of having a DM component with vanishing bulk viscosity: this is, $U\neq0,V=W=0$. Nevertheless, this fixed point might be achieved through another trajectory in phase space, as we shall see. While $V=0$ might limit the phase space availability, the new compelling properties are located inside the available part of the phase space according to the new phase space variables $(U,V,W)$. So, our approach is still quite general.

Note that the constraint equation is automatically satisfied by any solution of the above system, as taking its derivative and using the continuity equations for the fluid components leads directly to the Raychaudhuri equation (\ref{Raychaudhuri}). The EoS for the viscous fluid is, by definition, given by 
\begin{equation}
w_{\Pi}:= \frac{\Pi}{\rho_{m}}=V,\label{eff_EoS_mat}
\end{equation}
while the effective EoS is defined as
\begin{equation}
w_{\rm eff}:= - 1 -\frac{2}{3}\frac{H^\prime}{H^2} = \frac{1}{3} \; \left[ 1- (4 - 3 \gamma ) \frac{W}{V} + 3 \; W \right]. \label{eff_EoS_par}
\end{equation}
Both quantities are determined by the bulk viscosity pressure through the definition Eq.~(\ref{Phase_Space_variables}). Contrary to the unified models based on the Eckar's theory, $w_{\rm eff}$ does not depend only on the bulk viscosity coefficient, but on a combination with the other parameters, i.e, $k$ and $v$, in a non-linear way. This relaxes the usual condition $\xi_{0}\sim\mathcal{O}(1)$ required in Eckart’s theory for achieving a bulk-viscosity–driven acceleration, thereby providing a more natural mechanism for driving the current cosmic expansion. Similarly, the deceleration parameter can be expressed as
\begin{equation}
q := - \frac{\ddot{a} / a} {H^2} =  - 1 - \frac{H^\prime}{H} = 1- \frac{1}{2} (4 - 3 \gamma ) \; \frac{W}{V} + \frac{3}{2}\; W. 
\label{decel_par}
\end{equation}
Having established the dynamical equations for the full causal viscous scenario, we now proceed with a detailed analysis of the asymptotic behavior near the fixed points of the background dynamics. The strength of our approach lies in its applicability to arbitrary values of the viscous exponent $s$  and to the generalized Israel–Stewart evolution equation for the bulk viscosity, Eq.~(\ref{IS_modified}). Although the dynamical systems analysis is carried out for a general barotropic index $\gamma$—a choice that offers greater generality and flexibility for future observational constraints—there are parts of the discussion in which we must set $\gamma = 1$ to properly interpret the cosmological implications within unified dark matter scenarios.


\subsection{Analytical study of the fixed points structure for $s \neq 1/2$}
\label{sec:fixed_points}
We begin by determining the fixed points of the dynamical system defined by Eqs.\;(\ref{evol_U})-(\ref{evol_W}), followed by a linear stability analysis. To explore stability beyond the linear regime, we apply the general stability criterion formulated by Malkin \cite{Malkin_52}. The case $s \neq 1/2$ will be addressed first, while the particular value $s = 1/2$ due to its distinct mathematical structure, will be examined separately in subsections \ref{subsec:fixed_points_s_onehalf} and \ref{subsec:stability_s_half}.

The fixed points fall into the following three classes:

\begin{itemize}
  
\item  \textit{Type $Ia$} :   

\begin{equation}
U_{Ia} = 3^{(s-1/2)} \frac{\xi_0}{v^2 \gamma} \; \left( v^2 - \frac{1}{2} \right) \; \left( 1 - \frac{k^2}{v^2} \right), \quad V_{Ia} = - \gamma = \; W_{Ia}, \quad \text{with} \quad w_{\rm eff} = -1 \quad \text{and} \quad q = -1. 
\label{FP_Ia} 
\end{equation}
For an arbitrary effective barotropic constant $\gamma$, this fixed point defines a one-parameter family that physically corresponds to a pure viscous DM universe. This follows from the condition $V = W $ which implies $3 H^2 = 8 \pi G_{N} \rho_m $. In light of the constraint in Eq.~(\ref{constraint}), this further requires the radiation energy density to vanish ($\rho_r = 0$). Moreover, this fixed point leads to a cosmic acceleration as $q=-1$. Interestingly, the dissipative pressure $\Pi_{\infty} = -3 H_{\infty}^2 / (8\pi G_{N})$ is responsible for the accelerated expansion, where the asymptotic Hubble parameter is given by  
\begin{equation}
H_{\infty} = \sqrt{ \frac{3}{8 \pi G_N}} \; \left[ \frac{\xi_0}{v^2 \gamma} \;\left( v^2 - \frac{1}{2} \right) \left( 1 - \frac{k^2}{v^2} \right) \right]^{1/(1-2s)}. 
\label{H_asymp} 
\end{equation}
In this asymptotic limit, the accelerated expansion is described by a de Sitter solution of the form

\begin{equation}
a(t) = a_0 \; \exp {\left( H_{\infty}\; t \right)}, 
\label{de_Sitter} 
\end{equation}
with $t$ being the cosmological time. Using the above asymptotic form for the bulk viscosity pressure, or equivalently, $ \Pi_{as} =  -\gamma \rho_{m} $, the non-negativity of the entropy production expressed by Eq.(\ref{entropy_production}) leads to the lower bound on $v$
\begin{equation}
v  \geq k. 
\label{bound_on_v} 
\end{equation}
Therefore, the entropy production is non-negative, and the solution $U_{I}$ of Eq.(\ref{FP_Ia}) is non-trivial 
provided the two conditions $ v > k$, and $v^2 > 1/2$ are satisfied, which follows from its analytical expression. The positivity condition on the effective entropy follows from the integration of Eq. (\ref{entropy_production}) for constant temperature, giving rise to the asymptotic expression
\begin{equation}
S_{\rm eff} = S_0 + \left( \frac{3 \gamma H_{\infty}^2}{4 n_0 T v^2}\right) \left(v^2 - \frac{1}{2} \right) \exp({3H_{\infty} t)}. 
\label{eff_entropy} 
\end{equation}
As in the former case, the above expression is non-negative provided $ v^2 \geq 1/2$. We therefore conclude that for $v$ lying in the interval $m < v^2 \leq 2 - \gamma $, with $m = Max \{ 1/2, k^2 \}  $, the viscous DM fixed point of Eq.(\ref{FP_Ia}) leads to an accelerated universe, and is fully consistent with the thermodynamical requirements ($\dot{S} \geq 0$ and $S_{\rm eff} \geq 0$).

\item  \textit{Type $Ib$} :   

\begin{equation}
U_{Ib} =  \frac{3^{s-1/2} \xi_0}{v^2 \gamma} \; \left( \frac{V_{Ib}^2}{2 \gamma^2 v^2 } - 1 \right) \left( \frac{\gamma v^2}{V_{Ib}} + k^2 \right), ~ V_{Ib} = - \gamma + \frac{2/3}{1-s}, ~ \text{and}  \; W_{Ib} = 0; ~\text{with} ~w_{\rm eff} = \frac{1}{3} ~\text{and} ~ q = 1. 
\label{FP_Ib} 
\end{equation}
This fixed point corresponds in principle to a two-component fluid composed of DM and radiation, but the EoS coefficient $\omega_{eff} =1/3$ shows that the dominant component behaves as radiation, associated to very early times. Consequently, 
$H \rightarrow \infty$ and simultaneously $\Pi \rightarrow 0$, but remaining both ratios $\Pi/\rho_{m}$ and $\rho_{m}^{(1-s)} / H $  finite. As a consequence, the bulk viscosity pressure is also finite for $s\neq1$ and the universe undergoes a decelerated expansion. Note also that choosing $s = 0$ and $\gamma=1$ lead to a dark energy–like component with an effective EoS parameter $\omega_{m} = V_{Ib} = -1/3$. Nevertheless, this contribution is subdominant and, therefore, cannot account for early accelerated expansion. Most importantly, $s$ can not be chosen arbitrarily, but it must be restricted to 
the existence of the fixed point. Since $U$ by definition is a non-negative quantity, the dissipation exponent $s$ must lye within the interval 
\begin{equation}
1-\frac{2}{3\gamma (1+\sqrt{2}v)} < s < 1,
\label{bounds_Ib}
\end{equation}
which, for the limit case $v = \sqrt{2}$, goes into the condition $1 -1/3\gamma < s < 1$. For the case of interest $\gamma = 1$, the choice $s = 0$ is of course not admissible, and as $s$ approaches its lower limit, the dissipative fluid behaves like a stiff fluid with $w_{m}=V_{Ib} \to 1$. 

\item  \textit{Type $II$ }:

\begin{equation}
U_{II} = 0, \quad V_{II} = \pm \sqrt{2} \; \gamma \; v, \quad W_{II} = 0 , \quad \text{with} \quad w_{\rm eff} = \frac{1}{3} \quad \text{and} \quad q = 1.
\label{FP_II}
\end{equation}
This fixed point is compatible with a radiation dominated era, which follows from the value of the effective EoS parameter $\omega_{\rm eff} = 1/3$. It corresponds to the very early-time limit when $H \rightarrow \infty$, $\Pi \rightarrow 0$, and $\rho_{m} \rightarrow 0$; nevertheless, this behavior gives rise to a finite ratio $\Pi / \rho_{m}\to \pm \sqrt{2}\gamma v$. Consequently, no source for an accelerated expansion exists and $q$ necessarily assume a positive value.\\

\item  \textit{Type $III$ }:

\begin{equation}
U_{III} = 0, \; V_{III} = \pm \sqrt{2} \; \gamma \; v = W_{III}, \quad \text{with} \quad w_{\rm eff} = -1 + \gamma \left( 1 \pm \sqrt{2} v \right) \quad \text{and} \quad q = -1 + \frac{3}{2} \gamma \left( 1 \pm \sqrt{2} v \right).
\label{FP_II_b}
\end{equation}
This fixed point encompasses notable scenarios. To simplify the analysis, we focus on the case $\gamma = 1$. From the equality $V_{III} =  W_{III}$, follows that\footnote{Notice that $\Omega_{m} = W / V$ may become ill-defined as $V$ approaches zero in certain regions of the phase space. However, when both $W$ and $V$ simultaneously approach zero, the limit can be evaluated, yielding a finite and well-defined quantity. In this regime, $\Omega_{m}$ serves as a useful diagnostic for assessing the effective contribution of the viscous matter component to the total energy budget of the system.} $\Omega_{m} =1$, meaning that this fixed point is associated with pure DM. Moreover, for the upper sign ($+$), the effective EoS parameter is consistent with a stiff fluid, since the lower bound on the dissipative contribution to the sound speed, $v^2 > 1/2$, leads to the condition 

\begin{equation}
 w_{\rm eff} = \sqrt{2} ~v > 1 \quad \text{and} \quad q = \frac{1}{2} (1 + 3 \sqrt{2} ~ v) > 2,
\label{FP_III_stiff}
\end{equation}
while the lower sign ($-$) corresponds rather to a phantom DM fluid:
\begin{equation}
 w_{\rm eff} = -\sqrt{2} ~v < -1 \quad \text{and} \quad q = \frac{1}{2} (1 - 3 \sqrt{2} ~ v) < -1.
\label{FP_III_phantom}
\end{equation}
We now examine the potential emergence of singularities in the scale factor $a$, and consequently in the Hubble parameter $H$, for the last case. From the definition of the phase-space variables Eq.~(\ref{Phase_Space_variables}) (see also Eq.~(\ref{old_Phase_Space_variables})), it follows that for $U = 0$ and $\Omega_{m} = 1$, the relation $U \equiv \Omega_{m}^{1-s} X^{1/2 - s}$ implies that $(8 \pi G_{N} H^2)^{1/2 - s}$ must vanish identically. For values of the viscous exponent $s > 1/2$, and under an effective phantom behavior of DM, i.e. $\omega_{eff} < -1$, corresponding to the lower sign choice in Eq.~(\ref{FP_II_b}), the fulfillment of the above conditions implies that the Hubble parameter must diverge at this fixed point. Such a divergence signals the onset of a Big Rip singularity \cite{Caldwell:1999ew}. We will analyze further this fixed point when computing its stability properties in the next subsection. 

\end{itemize}
The main properties of these fixed points, along with the conditions for their existence, are summarized in Table~\ref{tab1:fixed_points_s_general}, providing a concise overview of the relevant cosmological regimes.
\begin{table*}[htp]
\centering  
\caption{Fixed points of the autonomous system given by Eqs.~(\ref{evol_U}), (\ref{evol_V}), and (\ref{evol_W}) for different values of the bulk-viscosity exponent $s \neq 1/2$, together with the corresponding conditions for their existence. The main cosmological features associated with each fixed point are also listed. This setup reduces to the unified DM model when $\gamma = 1$.}
\begin{ruledtabular}
\begin{tabular}{ccccccccccc}
Exponent& Point & $\Omega_{r}$ & $\Omega_{m}$ & $w_{\Pi}$ & $w_{\rm eff}$ & \text{Existence} & \text{Acceleration}
  \\ \hline
  \multirow{3}{*}{$s\neq\frac{1}{2}$} &  
 $P_{Ia}$ & $0$ & $1$ & $-\gamma$ & $-1$ &  $v^{2}>\frac{1}{2}$, $v > k$ & $\text{Yes}$\\
 & $P_{Ib}$  & $0$ & $1$ & $- \gamma + \frac{2/3}{1-s}$ & $\frac{1}{3}$ &  $1-\frac{2}{3\gamma (1+\sqrt{2}v)} < s < 1$ & $\text{No}$\\
 & $P_{II}^{+}$ & $1$  & $0$ & $\sqrt{2} \gamma v$ & $\frac{1}{3}$ & $s<1$, $v^2 > \frac{1}{2}$ & $\text{No}$\\
  & $P_{II}^{-}$  & $1$ & $0$ & $-\sqrt{2} \gamma v$ & $\frac{1}{3}$ &  $s<1$, $v^2 > \frac{1}{2}$ & $\text{No}$\\
 & $P_{III}^{+}$&$0$ & $1$& $\sqrt{2} \gamma v$ & $-1 + \gamma \left( 1 + \sqrt{2} v \right)>1$ & $v^{2}>\frac{1}{2}$, $s < \frac{1}{2}$ & $\text{No}$\\
 & $P_{III}^{-}$&$0$ & $1$& $-\sqrt{2} \gamma v$ & $-1 + \gamma \left( 1 - \sqrt{2} v \right)<-1$ & $v^{2}>\frac{1}{2},\; s>\frac{1}{2}$ & $\text{Yes}$
 \\
 \end{tabular}
\end{ruledtabular}\label{tab1:fixed_points_s_general}
\end{table*}
\subsection{Stability analysis of the fixed points for  $s \neq 1/2$}
The stability properties of the fixed points are crucial for assessing whether the nonlinear viscous DM model can accurately reproduce the observed cosmic evolution, as supported by a wide range of cosmological observations. To this end, we perform a linear stability analysis—an effective method for probing the behavior of a dynamical system in the vicinity of its critical points. Specifically, we analyze the system defined by Eqs. (\ref{evol_U})–(\ref{evol_W}), which govern the evolution of the dynamical variables $(U, V, W)$ that span the phase space of this cosmological framework. We first compute the derivatives of the functions on the r.h.s. of the dynamical system to obtain the Jacobian matrix. Finally, from its characteristic eigenvalues we extract the stability properties of the fixed points. We have the following:
\begin{align}
F_{1,U}&= \left[ -3 (1-s) \left( \; \gamma + V \right) + 2 - (4- 3 \gamma) \frac{W}{2 V} + \frac{3}{2} \; W \right], \quad F_{1,V}= \left[ -3 (1-s) + (4- 3 \gamma) \frac{W}{ 2 V^2} \right] U \\
F_{1,W}&= \left[ 3 - (4- 3 \gamma) \frac{1}{ V} \right] \frac{U}{2} \\
F_{2,U}&= - 3^{3/2-s} \; \frac{v^2 \gamma}{\xi_{0}} V \left( 1 + \frac{k^2}{ v^2 \gamma} V \right)^{-1}, \quad F_{2,V}= - 3^{3/2-s} \; \frac{v^2 \gamma}{\xi_{0}} \; U \left( 1 + \frac{k^2}{ v^2 \gamma} V \right)^{-2} + \frac{3}{\gamma} V, \quad F_{2,W} = 0 \\
F_{3,U}&= - 3^{3/2-s} \; \frac{v^2 \gamma}{\xi_{0}} W \left( 1 + \frac{k^2}{ v^2 \gamma} V \right)^{-1} \\
F_{3,V}&= \left[ 3  \frac{v^2 \gamma}{V^2} + 3^{3/2-s} \; \frac{k^2}{\xi_{0}} U \left( 1 + \frac{k^2}{ v^2 \gamma} V \right)^{-2} + (4- 3 \gamma) \frac{W}{ V^2} - 3 (1- \frac{1}{2 \gamma}) \right] W \\
F_{3,W}&= -3 v^2 \gamma \left[ \frac{1}{V} + \frac{3^{1/2-s}}{\xi_{0}} \; U \left( 1 + \frac{k^2}{ v^2 \gamma} V \right)^{-1} \right]  + (4- 3 \gamma) + 2 W \; \left( 3 - \frac{(4- 3 \gamma) }{ V}\right) - 3 \; (1- \frac{1}{2 \gamma}) \; V 
\end{align}
Here, the comma denotes differentiation with respect to the scale variable $N = \ln a$, and $F_i$ (with $i = 1, 2, 3$) represents the right-hand side functions of the dynamical system defined by Eqs. (\ref{evol_U})–(\ref{evol_W}). \\

Now we proceed to compute the characteristic polynomial $ \det(\lambda I - F_{i,j}) = 0$, where $i \in {(1, 2, 3)} $ and $j \in (U, V, W)$ in the above expressions. By solving this characteristic equation, we can obtain the eigenvalues for each fixed point. Subsequently, the eigenvalues allow us to determine whether a fixed point behaves as an attractor, a repeller, or a saddle, thereby defining its dynamical character. 
Regarding the applicability of linear stability theory beyond this regime, it is essential to address a technical point. Malkin's nonlinear stability theorem \cite{Malkin_52} asserts that the conclusions drawn from linear stability analysis concerning a fixed point $(U_*, V_*, W_*)$ remain valid beyond the linear approximation provided that: i) the right-hand side functions of the dynamical system defined by Eqs. (\ref{evol_U})-(\ref{evol_W}), denoted as $F_i(U, V, W)$, satisfy the following condition in a sufficiently small neighborhood of the fixed point:
\begin{equation}
\left|F_i (U-U_*, V-V_*, W-W_*) \right| \leq \mathcal{N} \left(U^{2} + V^{2} + W^{2} \right)^{1/2 + \alpha},
\label{malkin}
\end{equation}
where $\left|F_i \right|$ stands for the absolute value of the function $F_i$, and $\mathcal{N}$ and $\alpha$ are positive constants; and ii) all real parts of the eigenvalues of the Jacobian’s characteristic equation are negative, the point remains as an attractor; or conversely, the presence of at least one eigenvalue with a positive real part renders the fixed point unstable. This mathematical aspect is crucial in order to extend the validity of our results to the nonlinear region of the dynamical system. Now, we delve into the stability properties of suitable fixed points associated with those exponent values for which both types of fixed points exist —radiation and viscous matter—leading to a decelerated and accelerated expansion, respectively. \\

Type ${Ia}$ : 

\begin{equation}
\quad U_{Ia} = 3^{(1/2-s)} \frac{\xi_0}{v^2 \gamma} \; \left( v^2 - \frac{1}{2} \right) \; \left( 1 - \frac{k^2}{v^2} \right), \quad V_{Ia} = -  \gamma = W_{Ia}, \quad \text{together with} \quad w_{\rm eff} = -1  \quad \text{and} \quad q = -1 .
\label{FP_Ia_Stab}
\end{equation}

In terms of the coefficients 

\begin{equation}
\alpha = \frac{v^2 \gamma^2}{\xi_{0}}, \quad A = \left( 1 - \frac{k^2}{ v^2 } \right)^{-1}, \quad B = ( v^2 - 1/2),   
\end{equation}
the Jacobian matrix has the form \\

\begin{centering}
 
\begin{equation}
\begin{matrix}
 \det (\lambda I - F_{i,j}) = \left[ \begin{array}{ccc}
\lambda & -U_{I} \left[ 3 (s - 1/2 )- \frac{2}{ \gamma} \right]  & -\frac{2}{\gamma} U_{I} \\
- 3^{(3/2 -s)} \alpha A & \lambda + 3 \; \left[ 1 + 3 A B \right] & 0 \\
- 3^{(3/2 -s)} \alpha A & -1 + 3 A B & \lambda + 4 \end{array} \right]
\end{matrix}
\label{Jacobian_Ia}
\end{equation}

\end{centering}
The associated eigenvalues are given by

\begin{equation}
\lambda_{1} = -4, \quad \lambda_{\pm} = -\frac{3}{2}( 1 + A B) \pm \sqrt{\frac{9}{4}( 1 + A B)^2 + 9 \gamma \; B \; (s - 1/2) }.
\label{eigenvalues_Ia}
\end{equation}
This fixed point should have only negative eigenvalues—characteristic of an attractor—since it represents an accelerated expansion ($q = -1$) driven by the viscous DM component. Moreover, it is expected to asymptotically approach a de Sitter solution (see Eq.~(\ref{de_Sitter})). As a consequence, and based on the above equations, this behavior is sustained only if $s < 1/2$, given that $A, B \geq 0$. These conditions stem from the requirements imposed by the second law of thermodynamics and the need to ensure the positivity of the effective entropy (see Eq.~(\ref{bound_on_v})). This constitutes a key outcome of the stability analysis, namely that only viscous exponents satisfying $s < 1/2$ are viable for bulk viscosity–driven accelerated scenarios, as this fixed point acts, in accordance with the cosmological observations, as an \textit{attractor}.\\

Type ${Ib}$ : 

\begin{equation}
\quad U_{Ib} = \frac{3^{s-1/2} \xi_{0}}{ v^2 \gamma} \; \left( \frac{V_{Ib}^2}{2 v^2 \gamma^2}-1 \right) \left( \frac{v^2 \gamma}{V_{Ib}} + k^2 \right), ~ V_{Ib} = - \gamma + \frac{2/3}{1-s}, ~ W_{Ib} = 0; ~ \text{with} ~ w_{\rm eff} = \frac{1}{3} ~\text{and} ~ q = 1 .
\label{FP_Ib_Stab}
\end{equation}
The Jacobian matrix has the form 

\begin{centering}
 
\begin{equation}
\begin{matrix}
 \det (\lambda I - F_{i,j}) = \left[ \begin{array}{ccc}
\lambda & -F_{1,V}  & -F_{1,W} \\
- F_{2,U} & \lambda - F_{2,V} & 0 \\
0 & 0 & \lambda - F_{3,W} \end{array} \right]
\end{matrix}
\label{Jacobian_Ib}
\end{equation}

\end{centering}
Solving the characteristic equation, we obtain the associated eigenvalues 

\begin{equation}
\lambda_{3} = F_{3,W}, \quad \lambda_{\pm} = -\frac{3}{2}( 1 + A B) \pm \sqrt{\frac{9}{4}( 1 + A B)^2 + 9 \gamma \; B \; (s - 1/2) }.
\label{eigenvalues_Ib}
\end{equation}\\
Since the general analytical expression for $\lambda_{3}$ is not particularly illuminating, we examined limiting cases for the model parameters and found that it remains always negative. In contrast to $\lambda_{\pm}$ given in Eq.~(\ref{eigenvalues_Ia}), here the parameter $s$ is restricted to the range $1 - \frac{1}{3}\gamma < s < 1$, which leads to $\lambda_{+}$ being positive while $\lambda_{-}$ remains negative. Therefore, this fixed point is identified as a \textit{saddle}. However, it cannot coexist together with the de Sitter solution, compromising thus its cosmological viability. \\

Type $II$ :
\begin{equation}
\quad U_{II} = 0 = W_{II}, \quad V_{II} = \pm \sqrt{2} \gamma v, \quad \text{together with} \quad w_{\rm eff} = \frac{1} {3}  \quad \text{and} \quad q = 1 .
\label{FP_II_Stab}
\end{equation}
Similarly, evaluating this fixed point yields the following eigenvalues:

\begin{equation}
\lambda_{1} = -3 \gamma (1-s) ( 1 \pm \sqrt{2} v), \quad \lambda_{2} = \pm 3 \sqrt{2} v, \quad \lambda_{3} = 4 - 3 \gamma (1 \pm \sqrt{2} v ).
\label{eigenvalues_II}
\end{equation}

As noted previously, this fixed point corresponds to a radiation-dominated era. Moreover, since no dissipative processes due to viscosity are present, the cosmic expansion decelerates over time, characterized by a deceleration parameter $q = 1$. To consistently reproduce the well-established cosmological evolution, this fixed point must act as a \textit{repeller}, or a \textit{saddle} point, meaning that at least one of the associated eigenvalues should be positive. Remarkably, this result holds irrespective of the specific value of the viscosity exponent $s$, as long as $s \neq 1$; otherwise, this direction would become marginal. To demonstrate this, we recall the thermodynamical constraint $v^2 > 1/2$ (see Eq.~(\ref{eff_entropy})), which implies $ (1-\sqrt{2} \; v )< 0 $. As a result, the eigenvalues $\lambda_{1}$ and $\lambda_{3}$ are positive (negative) for the lower (upper) sign choice $-$ ($+$), while $\lambda_{2}$ remains negative (positive). Hence, this fixed point is a \textit{saddle} point, consistent with the cosmological dynamics of the early universe. \\

Type $III$ :
\begin{equation}
U_{III} = 0, \quad V_{III} = \pm \sqrt{2} \gamma v = W_{III}, \quad \text{together with} \quad w_{\rm eff} =  -1 + \gamma ( 1 \pm \sqrt{2} v ) \quad \text{and} \quad q = -1 + \frac{3}{2} \gamma \left( 1 \pm \sqrt{2} v \right). 
\label{FP_III_Stab}
\end{equation}

As discussed previously, this fixed point describes a pure viscous DM fluid, as $V_{III} = W_{III}$ ($\Omega_{m} = 1$). The value $U_{III} = 0$ implies that $H^{(1-2s)}$ must vanish asymptotically, which occurs for $s > 1/2$ if the Hubble parameter diverges, while for $s < 1/2$, $H$ must vanish. To understand the physical underlying scenarios described by each choice of possible $s$-values, we compute the eigenvalues associated to this fixed point, obtaining

\begin{equation}
\lambda_{1} = 3 \gamma (s - 1/2) ( 1 \pm \sqrt{2} v), \quad \lambda_{2} = \pm 3 \sqrt{2} v, \quad \lambda_{3} = - 4 + 3 \gamma (1 \pm \sqrt{2} v ).
\label{eigenvalues_III}
\end{equation}

For the upper sign choice ($+$), the EoS parameter is consistent with a stiff fluid, $\omega_{\rm eff} \approx 1$. In addition, $q > 0$ implies that the universe undergoes a decelerated expansion. This fixed point behaves as a \textit{saddle} point when $s<1/2$, since $\lambda_{1}<0$, while both $\lambda_{2}$ and $\lambda_{3}$ are always positive.

For the alternative sign choice ($-$), the EoS parameter satisfies $\omega_{\rm eff} < -1$, consistent with a phantom-like behavior of the DM fluid. This remarkable result stems from the nonlinear description of viscosity and leads to an accelerated expansion, since $q < -1$. Moreover, consistency with the well-established evolution of the universe requires that this fixed point corresponds to an attractor, i.e. all eigenvalues must be negative. Since the eigenvalues $\lambda_{2}$ and $\lambda_{3}$ are already negative, the stability condition reduces to requiring $\lambda_{1} < 0$, which holds only for $s > 1/2$. Hence, this condition is consistent with the existence of a Big Rip singularity, as conjectured early. Indeed, since $U_{III} = 0$ implies that for $s > 1/2$ the Hubble parameter diverges ($H \rightarrow \infty$), the universe is driven toward an \textit{attractor} fixed point dominated by an effective phantom-like viscous DM component.\\

In summary, the first two fixed points discussed above correspond, respectively, to viscous DM and radiation-dominated eras, thus reproducing the expected expansion phases of the Universe, where viscosity becomes negligible during matter domination. This consistency is further validated numerically in the next section. 
The viscous DM dominated scenario can lead to viscosity-driven accelerated expansion of the de Sitter type for $s<1/2$. Additionally, there exists another attractor solution exhibiting phantom-like behavior for $s>1/2$, in which the effective EoS is ultimately determined by the non-linear effects of the theory. A further (positive) branch of this solution displays an effective stiff-fluid behavior at early times. Consequently, the system consistently reproduces a wide range of cosmologically relevant dynamics, including standard radiation domination and late-time acceleration, with the possibility of asymptotically approaching a phantom regime. More importantly, all these fixed points satisfy the expected stability properties, as summarized in Table~\ref{tab2::eigenvalues_s_general}.

\begin{table*}[htp]
\centering  
\caption{Eigenvalues and stability conditions for determining the dynamical character of each fixed point for the bulk viscous unified DM model with exponents $s \neq 1/2$.}
\begin{ruledtabular}
\begin{tabular}{ccccccccccc}
Exponent & Point & $\lambda_{1}, \lambda_{2},\lambda_{3}$  & \text{Stability} \\ \hline
 \multirow{2}{*}{$s\neq 1/2$} &
$P_{Ia}$ & Eq.~(\ref{eigenvalues_Ia}) &   \text{Attractor} if $s<\frac{1}{2}$ \\
 & $P_{Ib}$ & Eq.~(\ref{eigenvalues_Ib}) & $\text{Saddle} \;\text{if}\; 1 - \frac{1}{3}\gamma < s < 1$\\
 & $P_{II}^{+}$ & Eq.~(\ref{eigenvalues_II})  &$  \text{Saddle}\; s\neq1$ \\
 & $P_{II}^{-}$ & Eq.~(\ref{eigenvalues_II}) & $\text{Saddle}\; s\neq1$\\
 & $P_{III}^{+}$ & Eq.~(\ref{eigenvalues_III}) &  \text{Saddle} if $s<1/2$ \\
 & $P_{III}^{-}$ & Eq.~(\ref{eigenvalues_III}) & $\text{Attractor} \;\text{if}\; s>1/2$\\
\end{tabular}
\end{ruledtabular}\label{tab2::eigenvalues_s_general}
\end{table*}
%

\subsection{Analytical study of the fixed points structure for s = 1/2}
\label{subsec:fixed_points_s_onehalf}

For this case, the phase space variables $(U, V, W)$ defined in (\ref{Phase_Space_variables}) satisfy the relation $U^2 V = W$, and therefore become unsuitable to describe the dynamical system in this case. Moreover, the viscous DM fixed point given by Eq.~(\ref{FP_Ia}), which asymptotically corresponds to a de Sitter solution, ceases to exist in the limit $s \rightarrow 1/2$ (see Eq.(\ref{H_asymp}) for instance). To address this, we introduce a new set of variables that define a reduced phase space through the following relations:
\begin{align}
& Y \equiv \left(\frac{8\pi G_{N} \rho_{m}} {3 H^2} \right)^{1/2} \; \;\; V \equiv \frac{\Pi}{\rho_{m}},
\label{Phase_Space_variables_s_onehalf}
\end{align}
which leads to the following dynamical system of equations governing the cosmological evolution:  
\begin{equation}
Y^\prime =  \frac{Y} {2} \left[ \; 4 - 3\gamma - 3 V \right] (1-Y^2),
\label{evol_Y_s_onehalf}
\end{equation}
\begin{equation}
V^\prime = -3 v^2 \gamma \left[ \; 1 + \frac{1}{\xi_0} Y V  \; \left(1 + \frac{k^2}{v^2 \gamma} V \right)^{-1} \right] + \frac{3}{2 \gamma} V^2. 
\label{evol_V_s_onehalf}
\end{equation}
For this particular case, the deceleration parameter and the effective EoS parameter can be expressed in terms of the new phase-space variables as:
\begin{equation}
q 	= 1 + \left[ \; -2 + \frac{3}{2} ( \gamma + V ) \; \right] Y^2 , 
\label{decel_par_s_onehalf}
\end{equation}
and
\begin{equation}
w_{\rm eff}=  \frac{1}{3} \; \left( 2 q - 1 \right) 
\label{eff_EoS_par_onehalf},
\end{equation}
respectively. 

We begin by determining the fixed points of the dynamical system defined by Eqs.\;((\ref{evol_Y_s_onehalf})-(\ref{evol_V_s_onehalf})), followed by a linear stability analysis. Beyond the linear regime, we apply the general stability criterion formulated by Malkin\cite{Malkin_52}. The fixed points can be classified into the following three categories:

\begin{itemize}

\item  \textit{Type $I^{+}$}:   

\begin{equation}
Y_{I} = 1 \quad V_{I} = V_{I}^*, \quad \text{with} \quad q =  -1 + \frac{3}{2} ( \gamma + V^* ) \;  \quad \text{and} \quad  \;  w_{\rm eff} = \frac{1}{3} \; \left( 2 q - 1 \right).
\label{FP_I_s_onehalf} 
\end{equation}
This is characterized by the positive branch of $Y$, and $V_{I}^{*}$ corresponds to one real root of the cubic polynomial 
\begin{equation}
 P(V) = (V^2 - 2v^2 \gamma^2) (V + v^2 \gamma / k^2 ) - \frac{2 v^4 \gamma^3}{\xi_{0} k^2} V.
 \label{polyn_V_s_onehalf}
\end{equation}
Although analytical expressions for the roots of the above polynomial can be obtained, they are not particularly enlightening. Instead, we concentrate in this part on specific, physically motivated solutions and defer the discussion of the general case to the numerical analysis. A particularly notable case occurs when the bulk viscosity coefficient is given by the analytic expression:
\begin{equation}
    \xi_{0} = \gamma \; v^2 \left( v^2 - \frac{1}{2} \right)^{-1} \left( 1 - \frac{k^2}{v^2} \right)^{-1},\label{bulk_coeff_plus}
\end{equation}
For this choice of $\xi_{0}$, there exists a real analytical root $V_{I}^* = -\gamma$. This specific dependence of $\xi_{0}$ on the dissipative sound speed $v$, the barotropic EoS coefficient $\gamma$, and the nonlinear relaxation coefficient $k^2$ is consistent with the physical expectation that bulk viscosity $\xi$ should depend on internal fluid properties. In particular, this includes the energy density of the DM component itself, as captured in the power-law dependence expressed in Eq.~(\ref{power_law_viscosity}).

The other two roots of Eq.~(\ref{polyn_V_s_onehalf}) for the above choice of $\xi_{0}$ are given by

\begin{equation}
V_{\pm}^{+} = \frac{\gamma}{2}~\left(1 - \frac{v^2}{k^2}\right) \left[ 1 \pm \sqrt{1 + \frac{8}{k^2} \left(\frac{1}{v^2} - \frac{1}{k^2}\right)^{-2}}  \right],
 \label{roots_polyn_V_s_onehalf}
\end{equation}
which represent two additional non-trivial fixed points. To simplify the notation, we collectively label all these points as $P_{I}^{+}$, with subscripts $a, b$ and $c$ used to distinguish the three roots of the cubic equation. Specifically, we assign $P_{Ia}^{+}$ to $V = V_{+}^{+}$, $P_{Ib}^{+}$ to the case $V = -\gamma$, and $P_{Ic}^{+}$ to $V = V_{-}^{+}$. Notice that, for the case of interest $\gamma = 1$, $w_{\rm eff}$ coincides with $V$, which is equivalent to $w_{\Pi}$.

Notably, the fixed point $P_{Ib}^{+}$, corresponding to $V_{I}^{*} = -\gamma$, represents a purely viscous DM-dominated era when $\gamma = 1$, with an effective EoS entirely determined by the bulk viscous pressure, $w_{\Pi} = -1$. This suggests that the viscous component could effectively mimic dark energy, giving rise to the present accelerated expansion. Indeed, from the condition $Y = 1$ follows that $3 H^2 = 8 \pi G_{N} \rho_m $, which, given the constraint in Eq.~(\ref{constraint}), implies that the radiation energy density must vanish ($\rho_r = 0$).  Hence, the viscous DM component, initially described as a standard barotropic fluid, effectively generates a negative pressure contribution due to bulk viscosity. This dissipative pressure $\Pi_{as} = -3 \gamma H_{\infty}^2 / (8\pi G_{N})$ drives the accelerated expansion, corresponding to a de Sitter solution with a constant asymptotic Hubble parameter $ H_{\infty}$. However, unlike the general case with $s \neq 1/2 $, no explicit analytical expression for $ H_{\infty}$ can be obtained within this scenario, as the corresponding expression to Eq.~(\ref{H_asymp}) goes into a trivial identity for $s = 1/2$.

Using the above asymptotic form for the bulk viscosity, or equivalently, $ \Pi_{as} =  -\gamma \rho_{m} $, the non-negativity of the entropy production expressed by Eq.~(\ref{entropy_production}) leads to the following lower bound on $v$

\begin{equation}
v  > k. 
\label{bound_on_v_s_onehalf} 
\end{equation}
Additionally, the existence of the non-trivial solution $Y_{Ib} = 1$ and $V_{Ib} = - \gamma$ is ensured only if the condition $v^2 > 1/2$ is satisfied. This condition also guarantees that the viscous coefficient $\xi_{0}$ remains positive. The positivity condition on the effective entropy follows from the integration of Eq. (\ref{entropy_production}) for constant temperature, giving rise to the asymptotic expression
\begin{equation}
S_{\rm eff} = S_0 + \left( \frac{3 \gamma H_{\infty}^2}{4 n_0 T v^2}\right) \left(v^2 - \frac{1}{2} \right) \exp({3H_{\infty} t)}. 
\label{eff_entropy_s_onehalf} 
\end{equation}
As for the former condition, the above expression remains non-negative provided $ v^2 \geq 1/2$. 

We therefore conclude that for $v$ lying in the interval $m < v^2 \leq 2 - \gamma $, with $m = Max \{ 1/2, k^2 \}  $, the viscous DM fixed point $P_{Ib}^{+}$ leads to an accelerated universe expansion, and is fully consistent with the thermodynamical requirements ($\dot{S} \geq 0$ and $S_{\rm eff} \geq 0$). The same reasoning applies to the other fixed points, namely $P_{Ia}^{+}$ and $P_{Ic}^{+}$. Nevertheless, it is necessary to verify numerically for these fixed points that Eqs.~(\ref{entropy_production}) and (\ref{positive_entropy}) are satisfied, in order to ensure that the thermodynamic constraints are properly fulfilled. Within the chosen range of model parameters, one can verify that $P_{Ia}^{+}$ can also lead to accelerated expansion, exhibiting a wide range of effective EoS $w_{\rm eff}$ that encompass various dark energy behaviors — including quintessence-like dynamics, de Sitter expansion, and even phantom regimes. This diversity arises from the dependence of the viscous contribution on both $\xi_{0}$ and $v$, and will be further investigated through numerical analysis. In contrast, $P_{Ic}^{+}$ behaves as an effective stiff fluid during an early phase of the universe’s evolution. Such versatility highlights the rich phenomenology of cosmological models with bulk viscosity, where a single fluid component can interpolate between different cosmic acceleration scenarios.

\item  \textit{Type $I^{-}$}:

\begin{equation}
Y_{I} = -1 \quad V_{I} = V_{I}^*, \quad \text{with} \quad q =  -1 + \frac{3}{2} ( \gamma + V^* ) \;  \quad \text{and} \quad  \;  w_{\rm eff} = \frac{1}{3} \; \left( 2 q - 1 \right),
\label{FP_I_minus_s_onehalf} 
\end{equation}
where $Y$ takes the remaining sign, and $V_{I}^{*}$ denotes a real root of the cubic polynomial
\begin{equation}
 P(V) = (V^2 - 2v^2 \gamma^2) (V + v^2 \gamma / k^2 ) + \frac{2 v^4 \gamma^3}{\xi_{0} k^2} V.
 \label{polyn_V_s_Y_m_one_onehalf}
\end{equation}
Similarly to the former case, there is a physically interesting analytical solution for the particular choice 
\begin{equation}
    \xi_{0} = \gamma \; \left( 1 - \frac{1}{2~v^2} \right)^{-1} \left( 1 + \frac{k^2}{v^2} \right)^{-1}.\label{bulk_coeff_minus}
\end{equation}
In effect, the analytical root turns out to be $V_{I}^{*} = \gamma$. The remaining two roots of Eq.~(\ref{polyn_V_s_Y_m_one_onehalf}) for the above choice of $\xi_{0}$ are given by

\begin{equation}
V_{\pm}^{-} = -\frac{\gamma}{2}~\left( 1 + \frac{v^2}{k^2}\right) \left[ 1 \pm \sqrt{1 - \frac{8}{k^2} \left(\frac{1}{v^2} + \frac{1}{k^2}\right)^{-2}}  \right],
\label{roots_polyn_Y_mone_V_minus_s_onehalf}
\end{equation}
which represent two additional non-trivial fixed points. Moreover, the condition for the existence of these roots is that $8 k^2 \leq 1$ is satisfied, or, if $8 k^2 > 1$ holds, then $v^2 < k^2 /(\sqrt{8k^2}-1)$. As in the previous case, there exists two accelerated solution, given by $V = V_{\pm}^{-}$ but the stability of $V_{+}^{-}$ is compromised, as shown later. In particular, the solution with $V_{-}^{-}$ can span a wide range of values for the effective EoS, mimicking dynamics from quintessence to phantom-like behavior. This depends specifically on the values of $\xi_{0}$ and $k$, as will be investigated numerically. Last but not least, the analytical solution $V_{I}^{*} = \gamma$, with $\gamma = 1$, describes an effective stiff fluid during the early stage of the Universe's evolution. This reveals a rich cosmological scenario attributed to bulk viscosity. Following the previous notation, we associate $P_{Ia}^{-}$ with the solution $V = V_{+}^{-}$, $P_{Ib}^{-}$ with $V = V_{-}^{-}$, and $P_{Ic}^{-}$ with $V = \gamma$. For this case, with $\gamma = 1$, $w_{\rm eff}$ also coincides with $V$, which is equivalent to $w_{\Pi}$.

\item  \textit{Type $II$ }:

\begin{equation}
Y_{II} = 0, \quad V_{II} = \pm \sqrt{2} \gamma v, \quad \text{with} \quad q= 1, \quad \text{and} \quad w_{\rm eff} = \frac{1}{3}.
\label{FP_II_s_onehalf}
\end{equation}
This fixed point is compatible with a radiation-dominated era, which follows from the value of the effective EoS parameter $w_{\rm eff} = 1/3$, as well as from the condition $Y \propto (\rho_m)^{1/2} / H \rightarrow 0$. This limit must be taken with same care, as $V_{II} \equiv \Pi / \rho_{m} \neq 0$. Although both the numerator and denominator vanish in the limit, their ratio approaches the finite values $\pm \sqrt{2} \gamma v$,  indicating a fluid behavior that can be interpreted effectively either as stiff-like or phantom-like, depending on the sign. 
Hence, in this very-early-times limit $\rho_m $ vanishes, and hence, necessarily the dissipative bulk pressure must vanish as well.  As a consequence, no source for an accelerated expansion exists and $q$ necessarily assume a positive value.\\

Finally, for the sake of completeness we display the last fixed point of the dynamical system of Eqs.~((\ref{evol_Y_s_onehalf})-(\ref{evol_V_s_onehalf})). 

\item  \textit{Type $III$ }:

\begin{equation}
V_{III} = \frac{1}{3} (4 - 3 \gamma), \quad Y_{III} = \frac{\xi_{0}}{V_{III}} \left( \frac{V_{III}^2}{2 {\gamma}^2 v^2} - 1 \right) \left( \frac{k^2 }{ \gamma v^2} V_{III} + 1 \right), \quad \text{with} \quad q = 1 \quad \text{and} \quad w_{\rm eff} = \frac{1}{3}.
\label{FP_III_s_onehalf}
\end{equation}

As for the general case $s \neq 1/2$ (see fixed points of type {Ib}), this fixed point corresponds to a two-component fluid composed of DM and radiation. However, if we consider only the positive branch of the square root in the definition of $Y$, this variable must be non-negative. Consequently, the existence of this fixed point is restricted to the narrow condition $\gamma < 2/3$, which excludes the physically relevant case in which DM behaves as a nearly dust-like fluid—consistent with current cosmological observations. Therefore, this fixed point together with the positive branch in the definition of $Y$ should be discarded on physical grounds. Nevertheless, as the physical quantity is related to $Y^2$, one may use the negative branch as well, which leads to a genuine fixed point. It represents an exotic fluid whose EoS corresponds to radiation expanding without acceleration. Its behavior should be better described numerically.  

All these fixed points, including the analytical solution that imposes the stringent condition on $\xi_{0}$ given by Eq.~(\ref{bulk_coeff_plus}), are depicted in the phase space diagram in the right panel of Figure~\ref{fig:phase_space__shalf}, while the left panel illustrates the general case, obtained by numerically solving the cubic equation with $\xi_{0}$ treated as a free parameter. A detailed physical discussion of the stability properties is deferred to the numerical analysis section as well as some physical implications. Nevertheless, we summarize in Table \ref{tab2:fixed_points_s_half} the main cosmological properties of the fixed points, along with the conditions for their existence and acceleration.

\end{itemize}

\begin{table*}[htp]
\centering  
\caption{Fixed points of the autonomous system given by Eqs.~(\ref{evol_Y_s_onehalf}) and (\ref{evol_V_s_onehalf}) for a bulk-viscosity exponent $s=1/2$, together with the corresponding conditions for their existence. The main cosmological features associated with each fixed point are also listed. This setup reduces to the unified DM model when $\gamma = 1$. Here we present the particular solutions for $P_{Ib}^{+}$ and $P_{Ic}^{-}$, which correspond, respectively, to the de Sitter solution and a stiff fluid.}
\begin{ruledtabular}
\begin{tabular}{ccccccccccc}
Exponent& Point & $\Omega_{r}$ & $\Omega_{m}$ & $w_{\Pi}$ & $w_{\rm \rm eff}$ & \text{Existence} & \text{Acceleration}
  \\ \hline
  \multirow{4}{*}{$s=\frac{1}{2}$} &  
 $P_{Ia}^{+}$ & $0$ & $1$ & $+$ sign, Eq.~(\ref{roots_polyn_V_s_onehalf}) & $ -1 +\gamma + V_{+}$ &  Eqs.(\ref{entropy_production}) \text{and} (\ref{positive_entropy}) & \text{Yes,} \;$V_{+} < 2/3 -\gamma$\\
 & $P_{Ib}^{+}$ & $0$ & $1$ &  $-\gamma$ & $-\gamma$ &  $ \hat{\xi}_{0}\; \text{from}\; \text{Eq.}~(\ref{bulk_coeff_plus}),\; \text{and}\; v^{2}>\frac{1}{2}, k^{2}$ & $\text{Yes}$\\
  & $P_{Ic}^{+}$ &$0$
 & $1$ & $-$ sign, Eq.~(\ref{roots_polyn_V_s_onehalf}) & $ -1 +\gamma + V_{-}$ & Eqs.(\ref{entropy_production}) \text{and} (\ref{positive_entropy}) & $\text{No}$\\
 &  
 $P_{Ia}^{-}$ & $0$ & $1$ & $+$ sign, Eq.~(\ref{roots_polyn_Y_mone_V_minus_s_onehalf})& $ -1 +\gamma + V_{+}$ &  Eqs.(\ref{entropy_production}) \text{and} (\ref{positive_entropy}) & \text{Yes,} \;$V_{+} < 2/3 -\gamma$\\
 & $P_{Ib}^{-}$ & $0$ & $1$ & $-$ sign, Eq.~(\ref{roots_polyn_Y_mone_V_minus_s_onehalf})& $ -1 +\gamma + V_{-}$ &  Eqs.(\ref{entropy_production}) \text{and} (\ref{positive_entropy}) & \text{Yes,} \;$V_{-} < 2/3 -\gamma$\\
  & $P_{Ic}^{-}$ &$0$
 & $1$ & $\gamma$ & $2\gamma-1$ & $ \hat{\xi}_{0}\; \text{from}\; \text{Eq.}~(\ref{bulk_coeff_minus}),\; \text{and}\; v^{2}>\frac{1}{2} $ & $\text{No}$\\
 &  
 $P_{II}^{\pm}$ & $1$ & $0$ & $\pm \sqrt{{2}\gamma v}$ & $\frac{1}{3}$ & $v>\sqrt{2}\; k^{2}$ & $\text{No}$\\
 \end{tabular}
\end{ruledtabular}\label{tab2:fixed_points_s_half}
\end{table*}
%

\subsection{Stability of the fixed points for s = 1/2} \label{subsec:stability_s_half}
We now turn to the stability analysis of the fixed points to assess whether they can support a complete and observationally consistent dynamical evolution of the Universe. To this end, we begin by computing the derivatives of the functions that appear on the right-hand side of the dynamical system defined by Eqs.~(\ref{evol_Y_s_onehalf}) and (\ref{evol_V_s_onehalf}):
\begin{align}
F_{1,Y}&= \frac{1}{2} \left[~ 4 - 3 (\gamma + V ) \right]  \left( \; 1 - 3 Y^2 \right) , \quad F_{1,V}= - \frac{3}{ 2} Y ( 1 - Y^2 ) \\
F_{2,Y}&= - 3 \; \frac{v^2 \gamma}{\xi_{0}} V \left( 1 + \frac{k^2}{ v^2 \gamma} V \right)^{-1}, \quad  \; F_{2,V}= - 3 \; \frac{v^2 \gamma}{\xi_{0}} \; Y \left( 1 + \frac{k^2}{ v^2 \gamma} V \right)^{-2} + \frac{3}{\gamma} V .\label{jacobian_II}
\end{align}
Next, we perform the stability analysis by computing the characteristic eigenvalues of the Jacobian matrix evaluated at the fixed points of types ${I}$  and  ${II}$, in order to classify them as \textit{saddles}, \textit{repellers}, or \textit{attractors} according to standard linear stability theory.  To determine the stability properties of all types of fixed points, we generally need to solve the corresponding cubic equations numerically. For instance, in the case of type $I^{+}$, the fixed points are given by the roots of the cubic equation~(\ref{polyn_V_s_onehalf}), which admits three solutions. Among these, only one root can be obtained analytically, and even then only under specific conditions, as discussed earlier. 
Nevertheless, in certain cases it is possible to derive analytical stability conditions by exploiting mathematical simplifications. This approach allows us to generalize the results and obtain conditions that remain valid throughout the entire parameter space.

\begin{itemize}

\item  \textit{Type $I^{+}$} :

\begin{itemize}

\item \text{Point} $P_{Ia}^{+} :$ In the accelerated region defined by the condition $q < 0$, which explicitly requires $V_{+}^{+} < 2/3 - \gamma$, (see for instance Eq.~(\ref{FP_I_s_onehalf})), the effective EoS  for $\gamma = 1$ corresponds to a phantom regime with $w_{\rm eff} < -1$. For all cases with $Y = \pm 1$ (i.e., type $I^{\pm}$), the term $F_{1,V}$ vanishes. Consequently, the eigenvalues reduce to the diagonal entries of the Jacobian matrix. They are thus obtained directly from Eqs.(\ref{jacobian_II}), yielding 
%
\begin{equation}
    \lambda_{1} = -4 + 3 \gamma + 3 V_{+}^{+},\quad \lambda_{2} = - 3 \; \frac{v^2 \gamma}{\xi_{0}} \;  \left( 1 + \frac{k^2}{ v^2 \gamma} V_{+}^{+} \right)^{-2} + \frac{3}{\gamma} ~ V_{+}^{+}.\label{eigen_PIa}
\end{equation}
Both eigenvalues are negative, as verified as follows: i) the acceleration condition $V_{+} < 2/3 -\gamma$ ensures that $\lambda_{1}$ is negative; ii) the second eigenvalue, $\lambda_{2}$, is also negative, since for $\gamma = 1$, it is given by the sum of two negative contributions ($V_{+}^{+} < 0$). Hence, this fixed point corresponds to an \textit{attractor} (cf. point $P_{Ia}^{+}$ in Fig.~\ref{fig:phase_space__shalf}). This feature is particularly appealing for describing the late-time cosmic acceleration.\\
\item \text{Point} $P_{Ib}^{+} :$  This fixed point also describes an accelerated expansion phase, since $q = -1$. This acceleration is triggered by the viscous pressure $\Pi$. The eigenvalues of the linearized system around this point are given by
\begin{equation}
\lambda_{1} = -4, \quad \lambda_{2} = -3 \left[ 1 + \frac{v^2 \gamma}{\xi_{0}} \left( 1 - \frac{k^2}{v^2} \right)^{-2} \right].
\label{eigenvalues_Ib_s_onehalf}
\end{equation}
Both eigenvalues are negative throughout the entire physically viable parameter space, ensuring that $P_{Ib}^{+}$ is a stable \textit{attractor} (cf. point $P_{Ib}^{+}$ in Fig.~\ref{fig:phase_space__shalf}). Unlike $P_{Ia}^{+}$, where phantom behavior emerges, this solution corresponds to a viscous-driven acceleration that, under the specific condition given by Eq.~(\ref{bulk_coeff_plus}), leads to a de Sitter-type expansion with $w_{\rm eff} = -1$. This regime mimics a cosmological constant and represents an important limit of the model. However, when the parameters deviate from this particular condition, the solution exhibits a less extreme form of acceleration, with $-1 < w_{\rm eff} < -1/3$, corresponding to a quintessence-like behavior. Moreover, the dynamical evolution associated with this fixed point is consistent with the thermodynamic constraints $\dot{S} \geq 0$ and $S_{\rm eff} \geq 0$, provided that the model parameters $v$ and $k$ satisfy $m < v^2 \leq 2 - \gamma $, where $m = Max \{ 1/2, k^2 \}.$\\

\item \text{Point} $P_{Ic}^{+} :$ This fixed point is characterized by $V_{-}^{+} > 0$ (see Eq.(\ref{roots_polyn_V_s_onehalf})), which implies a positive viscous pressure $\Pi$. Consequently, there is no mechanism driving accelerated expansion, and necessarily $q > 0$. For $\gamma = 1$, the effective EoS parameter is $\omega_{\rm eff} = V_{-}^{+}$, whose explicit value follows from Eq.(\ref{roots_polyn_V_s_onehalf}) once the model parameters are specified. It takes positive values ranging from $0$ when $v^2 \gtrapprox k^2$ to $1$ in the limit $v \gg k$. This range suggests the presence of an exotic fluid, with an EoS parameter varying between dust and stiff matter. Explicitly, the eigenvalues are
\begin{equation}
    \lambda_{1} = -4 + 3 \gamma + 3 V_{-}^{+},\quad \lambda_{2} = - 3 \; \frac{v^2 \gamma}{\xi_{0}} \;  \left( 1 + \frac{k^2}{ v^2 \gamma} V_{-}^{+} \right)^{-2} + \frac{3}{\gamma}~ V_{-}^{+}.\label{eigen_PIc}
\end{equation}
The first eigenvalue, $\lambda_{1} = F_{1,Y}$, is always positive under the physical condition $v^2 > 1/2$. This condition ensures the positivity of the effective entropy expressed by Eq.~(\ref{eff_entropy_s_onehalf}). The second eigenvalue, $\lambda_{2} = F_{2,V}$, is also positive, provided that $\xi_{0}$ satisfies the following lower bound condition
\begin{equation}
 \xi_{0} > \frac{v^2~\gamma^2}{V_{-}^{+}} ~\left( 1 + \frac{k^2}{v^2} ~V_{-}^{+} \right)^{-2}. \label{eigenvalues_I_Plus_c_s_onehalf}
\end{equation}
As both eigenvalues are positive, this fixed point corresponds to a \textit{repeller} (cf. point $P_{Ic}^{+}$ in Fig.~\ref{fig:phase_space__shalf}). However, if the condition given by Eq.~(\ref{eigenvalues_I_Plus_c_s_onehalf}) is not satisfied, the fixed point becomes a \textit{saddle}. This rather exotic behavior of the effective fluid emerges during an early phase of the universe’s evolution. 

\end{itemize}

\item  \textit{Type $I^{-}$}:

\begin{itemize}

\item \text{Point} $P_{Ia}^{-} :$ This fixed point is characterized by $V_{+}^{-} < 0$ (see Eq.(\ref{roots_polyn_Y_mone_V_minus_s_onehalf})), which yields a negative viscous pressure $\Pi < 0$. This effective bulk pressure drives an accelerated expansion, since the condition $V_{+}^{-} < 2/3 - \gamma$ is always fulfilled, as it can be seen straightforwardly from its analytic expression.
It follows, that the effective EoS parameter becomes $\omega_{\rm eff} = V_{+}^{-} < -1/3$. Its specific value can be obtained from Eq.(\ref{roots_polyn_Y_mone_V_minus_s_onehalf}), once the model parameters are inserted in that expression. The value of the EoS parameter ranges from $-(1+\sqrt{1 - 16/k^4})$, when $v^2 \gtrapprox k^2$, to values $-(1+\sqrt{1 - 8 k^2})v^2/2k^2 < $ in the limit $v \gg k$. This range suggests the presence of viscous dark matter, with an EoS parameter varying between a phantom to dark energy fluid. Concerning the stability properties of this fixed point, the associated eigenvalues can be expressed from Eq.~(\ref{jacobian_II}) as

\begin{equation}
    \lambda_{1} = -4 + 3 \left( \gamma + V_{+}^{-} \right),\quad \lambda_{2} = 3 \; \frac{v^2 \gamma}{\xi_{0}} \;  \left( 1 + \frac{k^2}{ v^2 \gamma} V_{+}^{-} \right)^{-2} + \frac{3}{\gamma} V_{+}^{-}.\label{eigen_PI_a_minus}
\end{equation}
For $\gamma=1$, $\lambda_{1}$ turns out to be negative since $V_{+}^{-} < 0$, while the second eigenvalue $\lambda_{2}$, happens to be positive, provided the following inequality is satisfied: 
\begin{equation}
 \left|~V_{+}^{-}\right | ~\left(1 + \frac{k^2}{v^2 } ~V_{+}^{-}\right)^2 < \left(v^2 - \frac{1}{2} \right) \left( 1+ \frac{k^2}{v^2}\right). \label{V_plus_minus}
\end{equation}
The above condition is satisfied in the accelerated region $V_{+}^{-} < 2/3 - \gamma$, for a dissipative contribution to the sound speed $v$ slightly above the lower value $v^2 > 1/2$, i.e. $v^2 > 1/2 + 2/27$, to ensure the positiveness of the effective entropy explained below Eq.~(\ref{eff_entropy}).
Therefore, we conclude that this fixed point is a \textit{saddle} and thus fails to satisfy the physical requirement of consistently describing the accelerated era of the universe, which is driven by viscous dark matter as the dominant component from the recent past into the future. \\

\item \text{Point} $P_{Ib}^{-} :$ For this fixed point, $V_{-}^{-}$ is negative in the complete range of the model parameters, which implies a negative viscous pressure $\Pi$. This negative pressure could trigger an accelerated expansion, provided the deceleration parameter $q = -1 + 3 (\gamma + V_{-}^{-})/2$ becomes negative, i.e., when $V_{-}^{-} < 2/3 - \gamma$, or explicitly for $\gamma = 1$, when $v^2 > (1+ \sqrt{1+ 16 k^2})/12$ holds. In addition, the effective EoS parameter is $\omega_{\rm eff} = V_{-}^{-} < -1/3$, whose explicit value follows from Eq.(\ref{roots_polyn_Y_mone_V_minus_s_onehalf}) once the model parameters are specified. The negative value of the EoS parameter ranges from $-1$, when $v^2 \gtrapprox k^2$ and $v^2 = 1/2$, to values less than $-1$ in the limit $v \gg k$ for relative large $k$-values, $k \approx 2 \sqrt2$ and $v^2 \gg 8 $. This range suggests the presence of viscous dark matter, with an EoS parameter varying between a phantom to dark energy fluid. Now, concerning the stability properties of this fixed point, the associated eigenvalues are

%
\begin{equation}
    \lambda_{1} = -4 + 3 \left( \gamma + V_{-}^{-} \right),\quad \lambda_{2} = 3 \; \frac{v^2 \gamma}{\xi_{0}} \;  \left( 1 + \frac{k^2}{ v^2 \gamma} V_{-}^{-} \right)^{-2} + \frac{3}{\gamma} V_{-}^{-}.\label{eigen_PI_b_minus}
\end{equation}
$\lambda_{1}$ is automatically negative for $V_{-}^{-} < 2/3 - \gamma$, which corresponds to the existence condition for an accelerated expansion. The second eigenvalue $\lambda_{2}$, is also negative. Indeed the condition $\lambda_{2} < 0$ translates into the inequality

\begin{equation}
 \left|V_{-}^{-}\right| \left(1 + \frac{k^2}{v^2 \gamma} ~V_{-}^{-}\right)^2 > \left(v^2 - \frac{1}{2} \right) \left( 1+ \frac{k^2}{v^2}\right), 
\label{V_minus_minus}
\end{equation}
This inequality is satisfied throughout the full range of model parameters, provided the constraint $v^2 > 1/2 $ holds (see discussion below Eq.~(\ref{eff_entropy_s_onehalf})). 
Therefore, this fixed point acts as an \textit{attractor}, where the viscous pressure is responsible for driving the late-time accelerated expansion. \\
\item \text{Point} $P_{Ic}^{-} :$

\begin{equation}
Y_{Ib} = -1 \quad V_{I}^{*} = \gamma, \quad \text{with} \quad q = - 1 + 3 \gamma \quad \text{and} \quad  \;  w_{\rm eff} = -1 + 2 \gamma.
\label{FP_I_s_onehalf_xi0} 
\end{equation}

As $F_{1,V} = 0$, the eigenvalues are given by the diagonal terms of the Jacobian matrix 
\begin{equation}
\lambda_{1} = -4 + 6\gamma, \quad \lambda_{2} = 3 \left[ 1 + \frac{v^2 \gamma}{\xi_{0}} \left( 1 + \frac{k^2}{v^2} \right)^{-2} \right].
\label{eigenvalues_Ic_s_onehalf}
\end{equation}
For the case of particular interest $\gamma \approx 1$, since both eigenvalues are positive, this fixed point is identified as a \textit{repeller} (cf. point $P_{Ic}^{+}$ in Fig.~\ref{fig:phase_space__shalf}). Accordingly, the deceleration parameter $q$ is positive, indicating that no accelerated expansion occurs, as the effective bulk pressure $w_{\Pi} = V \approx1$ remains positive. This clearly corresponds to an effective stiff fluid dominating the early stages of the cosmological evolution. Furthermore, the physical region defined by $m < v^2 \leq 2 - \gamma $, with $m = Max \{ 1/2, k^2 \} $ ensures consistency with the thermodynamic requirements ($\dot{S} \geq 0$ and $S_{\rm eff} \geq 0$). \\

\item  \textit{Type $II$ }: %
Similarly to the former computation, as $F_{1,V} = 0$ for $Y=0$, the eigenvalues for this fixed point are
\begin{equation}
\lambda_{1} = 2 - \frac{3}{2} \gamma ~ ( 1 \pm \sqrt{2} v ), \quad \lambda_{2} = \pm \sqrt{2} ~v.
\label{eigenvalues_II_s_half}
\end{equation}

The fixed point corresponds for both possible signs to a \textit{saddle} point, representing an early, radiation-dominated phase in the cosmic evolution. For instance, for $\gamma = 1$, choosing the upper sign ($+$), it follows that $\lambda_{1} < -1 $, while $\lambda_{2} > 1$. Similarly, for the lower minus sign, $\lambda_{1} > 0 $, while  $\lambda_{2} < 0$.  These behaviors hold across the entire physically allowed range of the parameter $v$. Notably, the effective viscous EoS results in a phantom-like fluid when the minus sign is chosen, whereas for the positive sign and $\gamma = 1$, the fluid behaves like standard matter. Consequently, no stable past attractor exists in this scenario.

\end{itemize}

In summary, the case $s = 1/2$, as well as the general scenario, exhibits a rich variety of cosmological behaviors driven by the bulk viscosity pressure, which plays a crucial and dynamic role throughout the universe's evolution. At early times, the viscous DM component can effectively mimic a stiff fluid. During intermediate epochs—and depending on the model parameters—it may approximate standard DM (though not exactly pressureless), as the viscous effects become nearly negligible. Remarkably, at late times, the bulk viscosity becomes dominant again, triggering cosmic acceleration in a way that spans a wide range of dark energy-like behaviors, including quintessence, de Sitter expansion, and even phantom regimes. These late-time dynamics are determined by the non-linear bulk viscosity parameters $k$ and $v$, as well as the coefficient $\xi_0$. All these accelerated solutions are classified as attractor points of the dynamical system, as demonstrated by the linear stability analysis (see Table~\ref{tab4::eigenvalues_s_half} for a concise summary). Importantly, the entire framework—rooted in a non-linear theory of bulk viscosity—is fully consistent with thermodynamic constraints, reinforcing its physical viability and offering a novel approach to modeling late-time cosmic acceleration.

\begin{table*}[htp]
\centering  
\caption{Eigenvalues and stability conditions for determining the dynamical character for the bulk viscous unified DM model with exponent $s = 1/2$.}
\begin{ruledtabular}
\begin{tabular}{ccccccccccc}
Exponent & Point & $\lambda_{1}$, $\lambda_{2}$ & \text{Stability} \\ \hline
 \multirow{2}{*}{$s= 1/2$} &
$P_{Ia}^{+}$ & Eq.~(\ref{eigen_PIa})  &  \text{Attractor if} $V_{+} < 2/3 -\gamma$ \\
 & $P_{Ib}^{+}$ & Eq.~(\ref{eigenvalues_Ib_s_onehalf}) & $\text{Attractor} \;\text{if}\; v^{2}>\frac{1}{2}$\\
 & $P_{Ic}^{+}$ & Eq.~(\ref{eigen_PIc}) &  \text{Repeller iff} Eq.~(\ref{eigenvalues_I_Plus_c_s_onehalf}); \text{else saddle} \\
  & 
  $P_{Ia}^{-}$ & Eq.~(\ref{eigen_PI_a_minus}) & $\text{Saddle} \;\; \forall \xi_{0},k,v$ \\
 & $P_{Ib}^{-}$ & Eq.~(\ref{eigen_PI_b_minus}) &  \text{Attractor} if $V_{-}^{-} < 2/3 - \gamma$ and Eq.~(\ref{V_minus_minus}) \\
 & $P_{Ic}^{-}$ & Eq.~(\ref{eigenvalues_Ic_s_onehalf}) & $\text{Repeller} \;\; \forall \xi_{0},k,v$\\
  &
  $P_{II}^{\pm}$ & Eq.~(\ref{eigenvalues_II_s_half}) & $\text{Saddle} \;\; \forall \xi_{0},k,v$\\
\end{tabular}
\end{ruledtabular}\label{tab4::eigenvalues_s_half}
\end{table*}

\end{itemize}
\section{Numerical evolution}\label{sec:IV}

In this section, we numerically explore the phase space to validate the dynamical properties of the fixed points previously obtained. This analysis also serves to evaluate the cosmological viability of the model, particularly its ability to reproduce the expected transition of cosmic evolution: from a radiation-dominated era to a matter-dominated phase, followed by late-time accelerated expansion. The latter is driven by the effective pressure arising from bulk viscosity. The numerical procedure is carried out separately for the two cases, $s\neq1/2$ and $s=1/2$, following the distinction made in the analytical analysis. As discussed earlier, although the phase space is not compact, compactification is unnecessary because the current choice of variables provides access to the physically relevant region. Therefore, our exploration remains sufficiently general. 

For all numerical integrations, we initialize the system deep in the radiation-dominated epoch and evolve it forward in time. Initial conditions are chosen through a careful trial-and-error procedure to ensure that the system passes through a realistic cosmological sequence and reaches a phase of late-time accelerated expansion consistent with observations. 
Our goal is to verify numerically that the existence of the radiation fixed point is independent of the values of $s$, in contrast to the Eckart theory. More importantly, we aim to test the existence of the attractor solutions by exploring different initial conditions.

\subsection{the case $s\neq 1/2$}

For illustrative purposes, we select three representative values of the bulk viscosity exponent within the range $s<1/2$, where the de Sitter solution is the unique late-time attractor. Specifically, we consider $s=-1/2$, $s=0$, and $s=0.4$, which excludes the existence of the fixed point\footnote{We remind that the fixed point $P_{Ib}$ exists only for the interval $2/3 < s < 1$, and therefore, cannot be continuously connected to the de Sitter solution.} $P_{Ib}$. In the following, we then turn our attention to the alternative attractor---identified with the phantom regime---which arises for $s>1/2$.  In accordance with thermodynamic constraints, we fix $v = 0.71$ and $k = 0.3$, as variations in these parameters do not significantly alter the phase space trajectories. The bulk viscosity coefficient $\xi_{0}$ is varied in the range $\mathcal{O}(10^{-3})<\xi_{0}<\mathcal{O}(10^{-1})$. This choice, together with the other parameter values, ensures that our numerical integration remains within the physically meaningful domain, satisfying the existence and acceleration conditions listed in Table~\ref{tab1:fixed_points_s_general}, as well as the stability properties summarized in Table~\ref{tab2::eigenvalues_s_general}.

In our approach, the dynamical nature of the fixed points depends on a non-trivial combination of the non-linear parameters $k$ and $v$. This contrasts with Eckart and IS-based models that do not satisfy the far-from-equilibrium condition, in which the bulk viscosity coefficient $\xi_{0}$ plays the determining role. As a result, once these non-linear parameters are fixed, variations in $\xi_{0}$ are expected to produce only minor deviations in the numerical trajectories and slight shifts in the position of the type I fixed points in phase space, without affecting their asymptotic behavior.

\begin{figure*}
\centering
\includegraphics[width=0.47\hsize,clip]{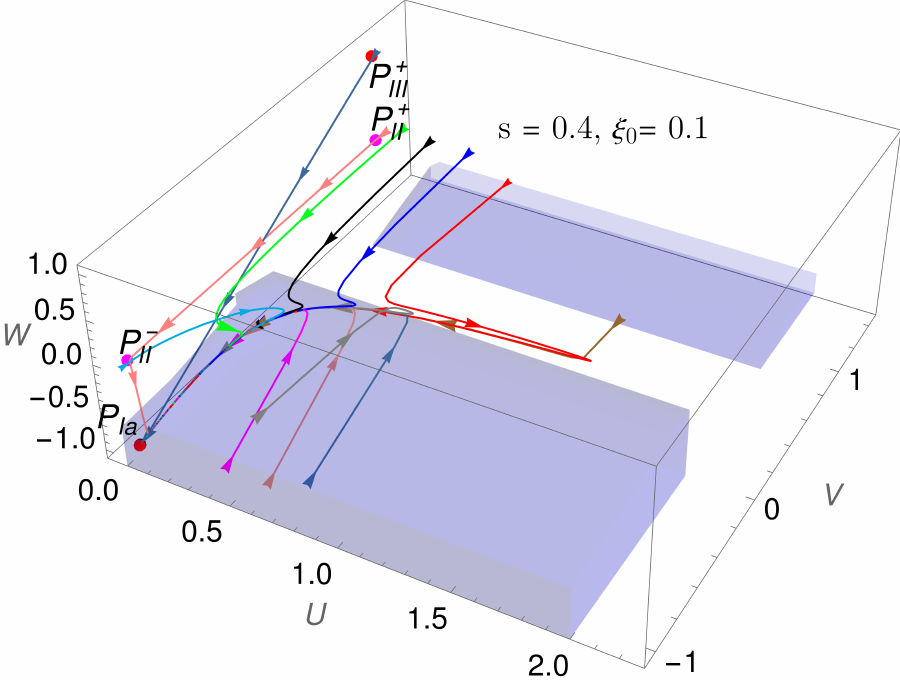}
\includegraphics[width=0.47\hsize,clip]{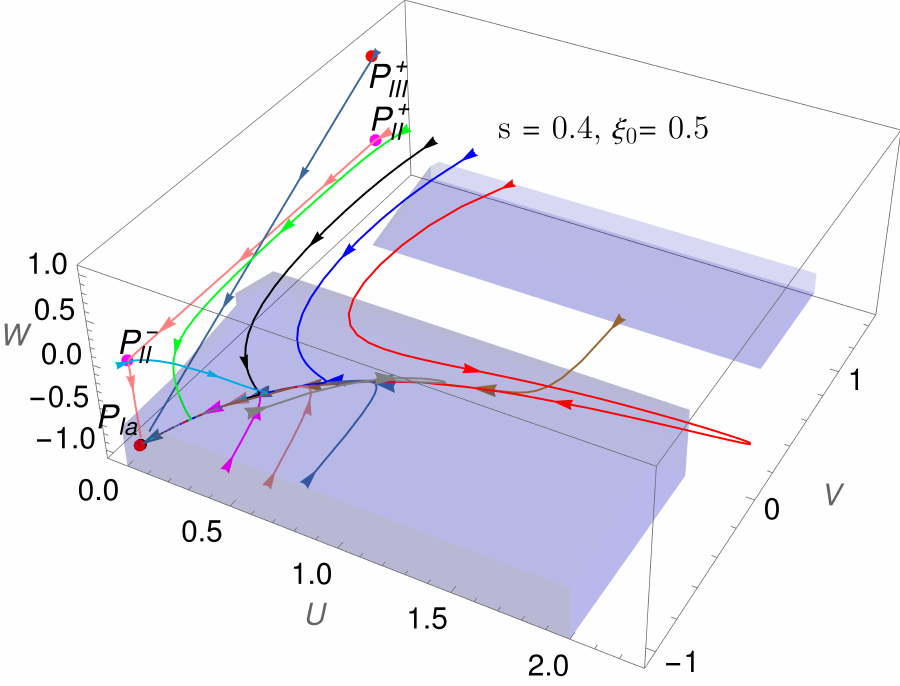}
\includegraphics[width=0.47\hsize,clip]{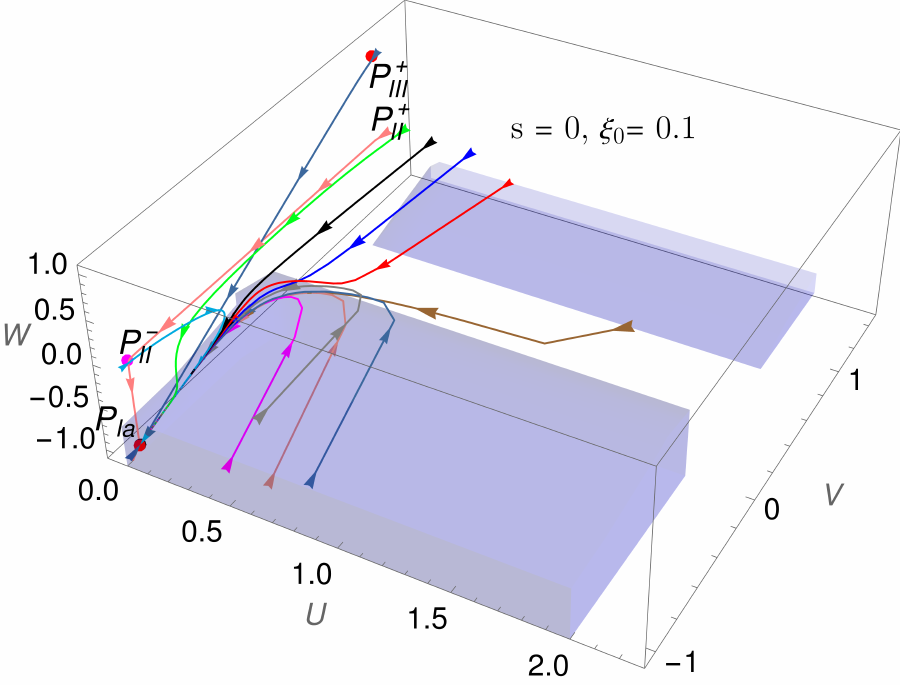}
\includegraphics[width=0.47\hsize,clip]{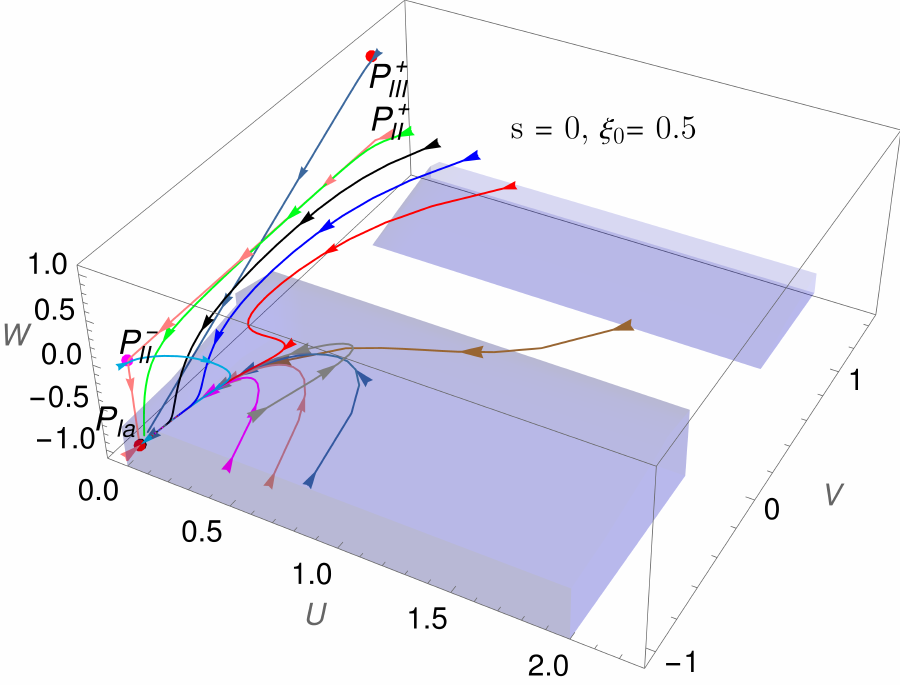}
\includegraphics[width=0.47\hsize,clip]{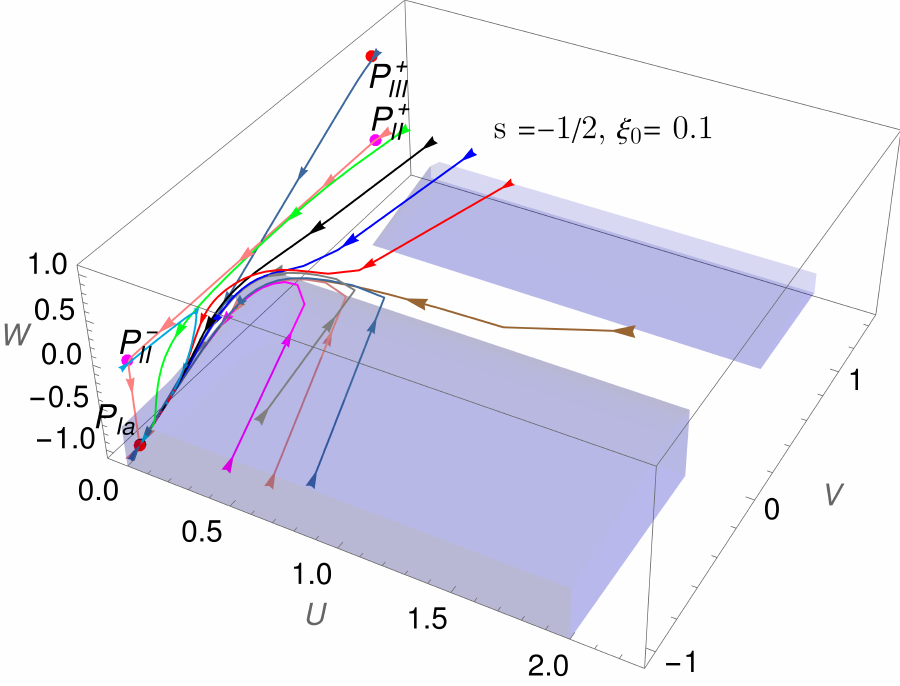}
\includegraphics[width=0.47\hsize,clip]{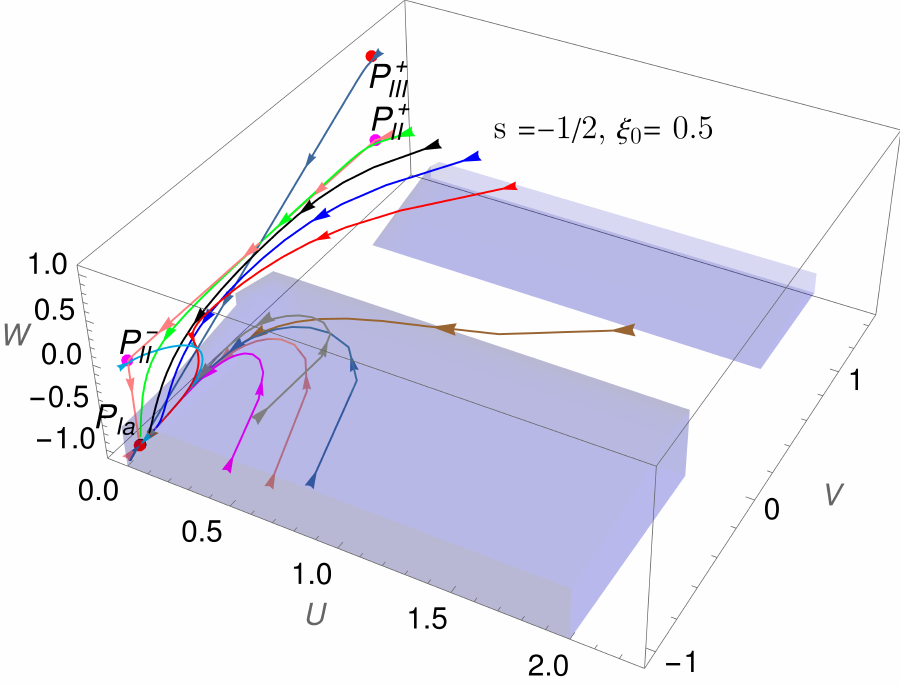}
\caption{Numerical trajectories in phase space with several initial conditions that capture the physical behavior of the dynamical system for $s\neq 1/2$. This evidences the attractor nature of the fixed point $P_{Ia}$. Initial conditions are taken close to the radiation point $P_{II}^{+}$. The latter is bridged to its negative branch $P_{II}^{-}$, as evidenced. Thus, trajectories starting strictly on $P_{II}^{+}$ will end up on $P_{II}^{-}$ and, subsequently, on the attractor point $P_{Ia}$. The point $P_{III}^{+}$ is a repeller, describing a stiff fluid at early times, while its negative branch, $P_{III}^{-}$, is a phantom fixed point. This cannot be visualized here due to the resolution scale and is instead represented in Figure \ref{fig:phase_space__sfree_phantom}. The point $P_{IV}$, corresponding to dark matter with $w_{\rm eff}=0$, is the only one that responds sensitively to changes in the bulk viscosity coefficient $\xi_{0}$. The blue region corresponds to the physical region where acceleration can take place, along with the condition $U > 0$.} \label{fig:phase_space__sfree}
\end{figure*}

All these features are illustrated in Figure~\ref{fig:phase_space__sfree}, which can be interpreted as follows. For a fixed value of $\xi_{0}$—specifically, $\xi_{0} = 0.1$ in the left panels and $\xi_{0} = 0.5$ in the right panels—we vary $s$ from a value close to its upper bound down to a negative reference value, $s = -1/2$, namely from the top to the bottom panels.
Comparing the three left panels (or the three right panels), one can observe that as $s$ decreases, the trajectories converge more rapidly along the attractor curve, leading to a stacking effect. This behavior is more pronounced for smaller values of $\xi_{0}$ (compare left versus right panels).

The light blue region indicates the physical domain of the phase space where cosmic acceleration occurs, while simultaneously satisfying the thermodynamic constraints. A general trend emerges: all trajectories tend to converge toward the de Sitter attractor point labeled $P_{Ia}$, regardless of the initial conditions. Some trajectories are initialized near the radiation-dominated point $P_{II}^{+}$, or even close to the stiff fluid point $P_{III}^{+}$, which characterizes the early universe. Interestingly, if the initial conditions are sufficiently close to $P_{II}^{+}$, the trajectory first passes through the radiation-like point $P_{II}^{-}$ before approaching the attractor $P_{Ia}$. This connection is illustrated by the pink trajectory. This demonstrates the various dynamical regimes that the viscous fluid can undergo during cosmic evolution, ultimately behaving as an effective dark energy component driving late-time acceleration.

\begin{figure*}
\centering
\includegraphics[width=0.47\hsize,clip]{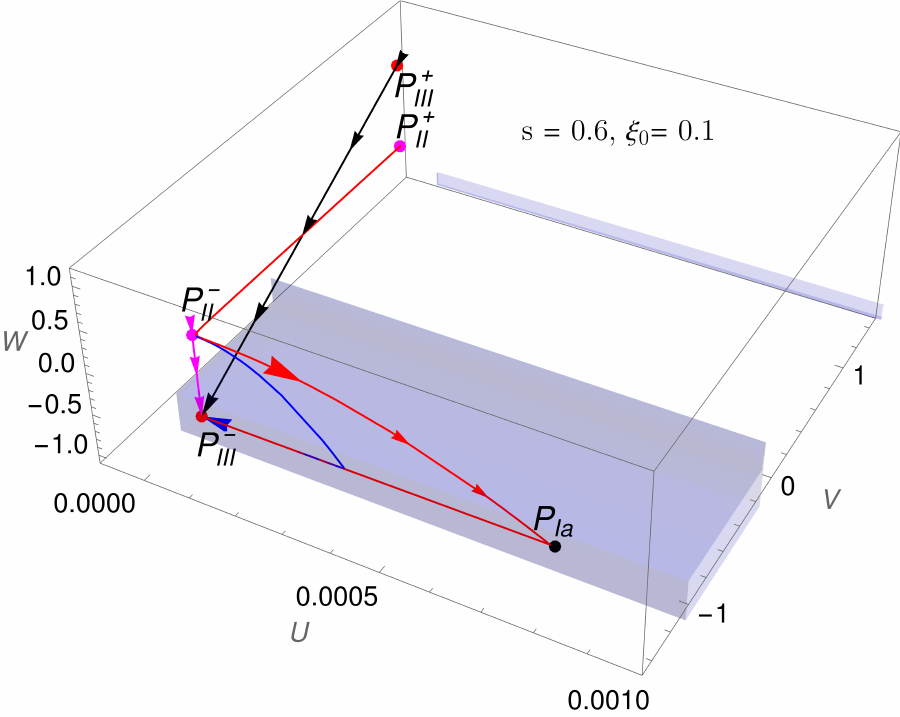}
\caption{Phase-space trajectories for the dynamical system with $s = 0.6$. Trajectories departing from $P_{II}^{-}$ evolve either toward the de Sitter (saddle) point $P_{Ia}$ (blue and red curves) or directly toward the effective phantom attractor $P_{III}^{-}$, characterized by $w_{\Pi} = -\sqrt{2} \, \gamma \, v \approx -1.00409$ with $v = 0.71$ (magenta and black curves), depending on the initial conditions. The basin-boundary saddle point $P_{II}^{-}$ separates these two intermediate paths. However, since $P_{Ia}$ is a saddle for $s > 1/2$, all trajectories ultimately converge to the phantom attractor, which is the only late-time stable solution. The blue-shaded region indicates the physically allowed domain with $U>0$, where cosmic acceleration can occur.} \label{fig:phase_space__sfree_phantom}
\end{figure*}

To illustrate the phantom attractor behavior associated with the fixed point $P_{III}^{-}$, we focus on the region $s > 1/2$. This point is located very close to the de Sitter solution. To highlight this, we zoom in on the $U$-axis near zero, as shown in Figure~\ref{fig:phase_space__sfree_phantom}. Initial conditions near the stiff fluid point $P_{III}^{+}$ (black curve) or the radiation-like point $P_{II}^{-}$ (magenta curve) lead the system directly toward the late-time phantom regime. A slight perturbation of initial conditions near $P_{II}^{-}$ (blue curve) reveals its saddle-type nature, deflecting trajectories toward the de Sitter point $P_{Ia}$ and effectively acting as a boundary between the basins of attraction of $P_{III}^{-}$ and $P_{Ia}$. We refer to this point as a \textit{basin-boundary saddle}. However, for $s > 1/2$, $P_{Ia}$ is no longer a true attractor but becomes a saddle, as evidenced by the numerical trajectory shown in red, which initially approaches $P_{Ia}$ but is eventually deflected toward the phantom attractor $P_{III}^{-}$.

\begin{figure*}
\centering
\includegraphics[width=0.47\hsize,clip]{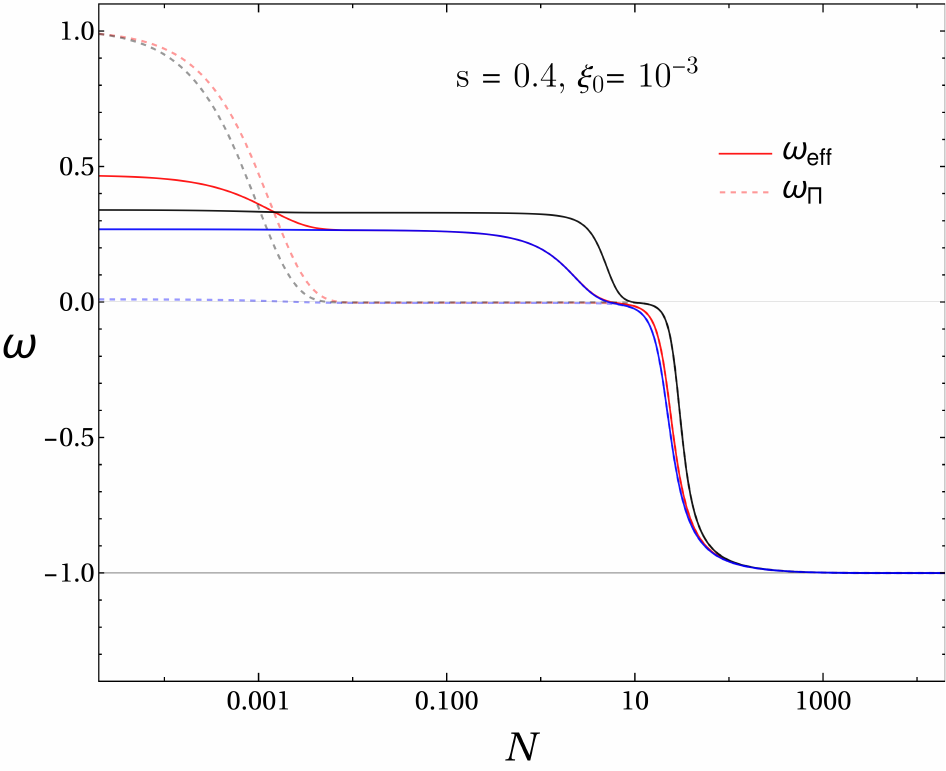}
\includegraphics[width=0.47\hsize,clip]{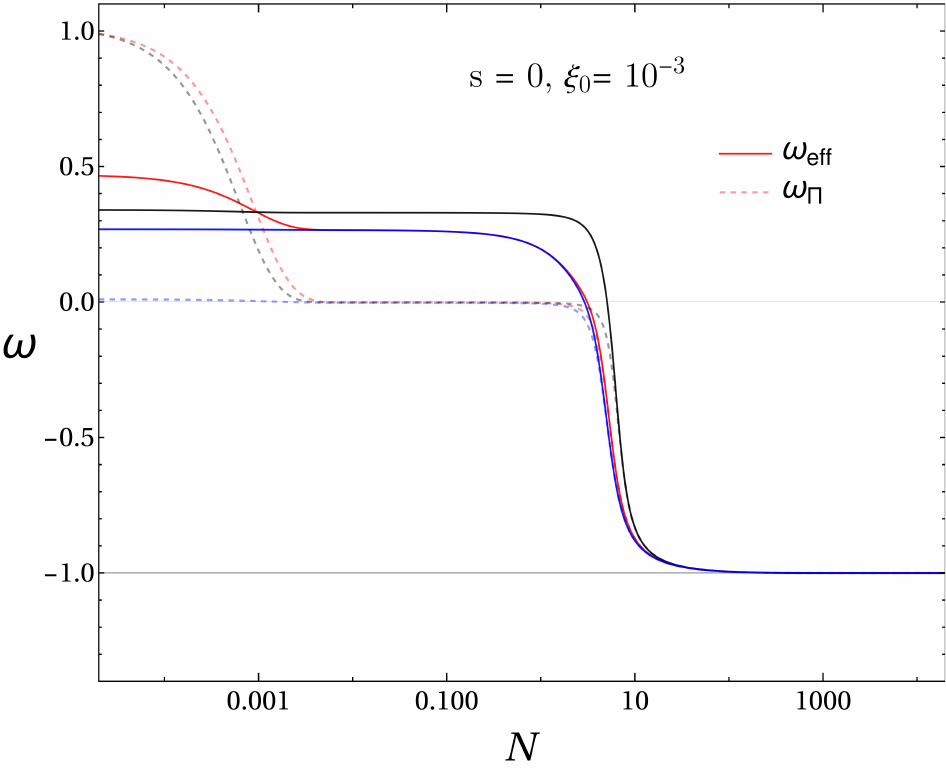}
\includegraphics[width=0.47\hsize,clip]{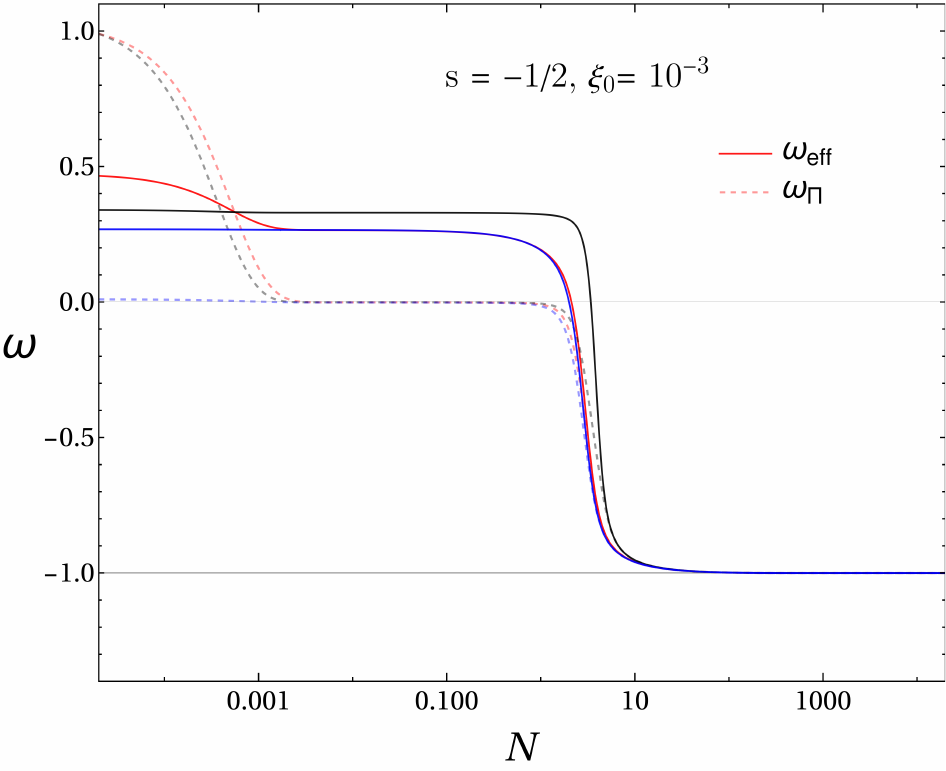}
\includegraphics[width=0.47\hsize,clip]{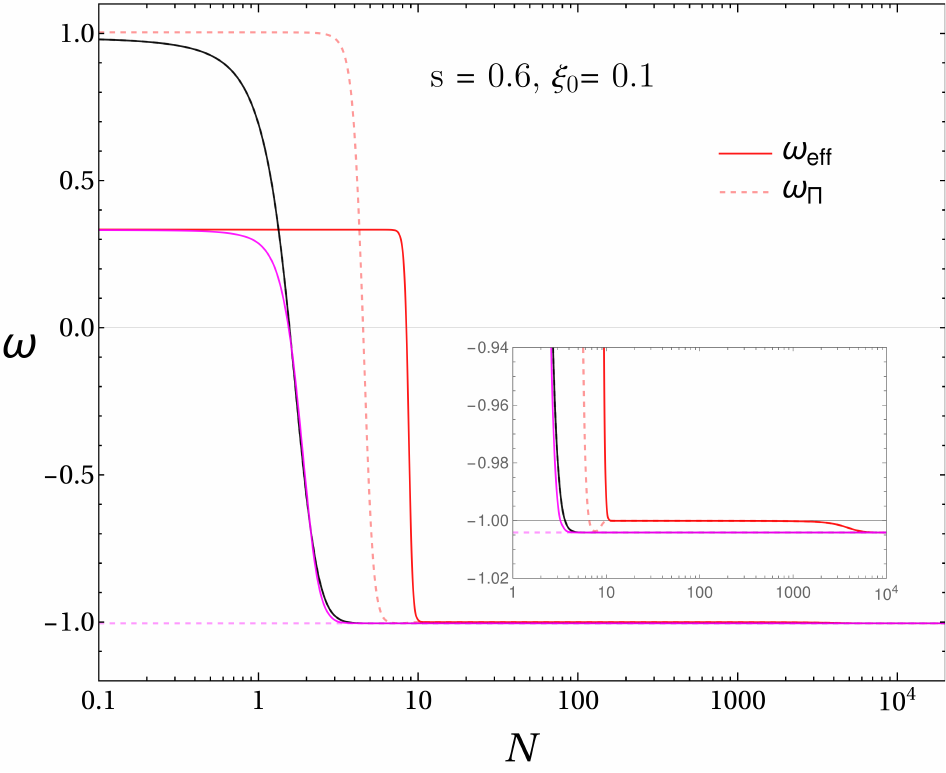}
\caption{Evolution of the viscous DM EoS parameter, $w_{\Pi}$, and the overall effective EoS, $w_{\rm eff}$, as a function of $N$. We adopt $\xi_{0} = 10^{-3}$, for the case $s<1/2$, while varying the viscous exponent $s$ to illustrate its impact on the global dynamics: $s = 0.4$ (top-left panel), $s = 0$ (top-right panel), and $s = -1/2$ (bottom-left panel). For these three cases, we have chosen the corresponding set of initial conditions: $U(t_{\rm init}) = 0.5$, $V(t_{\rm init}) = 0.99$,
$W(t_{\rm init})= 0.2$ for red curves; $U(t_{\rm init}) = 0.7$, $V(t_{\rm init}) = 0.99$, 
$W(t_{\rm init}) = 0.01$ for black curves; $U(t_{\rm init}) = 0.3$, $V(t_{\rm init}) = 0.01$, 
$W(t_{\rm init}) = 0.002$ for blue curves. Decreasing $s$ disrupts the smooth late-time evolution, hastening the onset of accelerated expansion. In the bottom-right panel, we consider $s > 1/2$ ($s=0.6$) with $\xi_{0} = 0.1$ to illustrate the phantom-like behavior, as highlighted in the inset (zoom-in) panel. For this plot, we have chosen, instead, the following set of initial conditions: $U(t_{\rm init}) = 10^{-6}$, $V(t_{\rm init}) = \sqrt{2} v$,
$W(t_{\rm init})= 0.001$ for red curves; $U(t_{\rm init}) = 0$, $V(t_{\rm init}) = 0.98$, 
$W(t_{\rm init}) = 0.98$ for black curves; $U(t_{\rm init}) = 0$, $V(t_{\rm init}) = -\sqrt{2}v$, 
$W(t_{\rm init}) = -0.001$ for magenta curves. In all plots we have adopted $v = 0.71$, and $k = 0.3$, as references values.} \label{fig:w_eff_sfree}
\end{figure*}

We conclude this section by examining the evolution of the viscous DM fluid—specifically, the equation-of-state parameter $w_{\Pi}$, which coincides with the effective DM EoS—together with the overall effective EoS $w_{\rm eff}$, in order to illustrate the global dynamics of the Universe. This is displayed in Figure \ref{fig:w_eff_sfree}. Unlike the previous numerical analysis of phase-space trajectories, here we adopt a deliberately small bulk viscosity coefficient, $\xi_{0} \sim 10^{-3}$, for the case $s<1/2$. This choice allows us to probe the analytical de Sitter solution, which is independent of $\xi_{0}$, and to highlight a distinctive feature compared to Eckart-based models, which require much larger values, $\xi_{0} \sim \mathcal{O}(1)$, to achieve similar late-time behavior within bulk viscous unified DM scenarios, where $w_{\rm eff}=-\xi_{0}$ \cite{Palma:2024qrw}. Therefore, adopting $s<1/2$ (see top and left bottom panels), all solutions converge to the de Sitter–like attractor $P_{Ia}$, characterized by $w_{\rm eff}=-1$.

In this non-linear causal framework, late-time accelerated expansion can be realized for extremely small values, while still providing a consistent matter-dominated era. Increasing $\xi_{0}$ shortens the duration of matter domination, whereas decreasing it prolongs that stage—without affecting the late-time attractor dynamics. Although we have investigated this feature in detail, here we focus on illustrating the impact of the viscous parameter $s$ on the global dynamics, fixing $\xi_{0} = 10^{-3}$, $v=0.71$ and $k=0.3$ for the sake of comparison.  We conclude that decreasing $s$ leads to a breakdown of smoothness in the late-time regime, effectively hastening the onset of accelerated expansion. This is illustrated in Figure \ref{fig:w_eff_sfree}. This effect cannot be compensated for by adjusting different initial conditions, but only by choosing an appropriate value of $\xi_{0}$ and the non-linear parameters $v$ and $k$. Such a choice constitutes a necessary condition for the model to satisfy, along with other cosmological constraints inferred from observations, in order to establish the cosmological viability of the scenario. A detailed investigation of these requirements lies beyond the main scope of this work, and is therefore left for future research.

As expected, for $s>1/2$, the late-time attractor is the phantom solution represented by $P_{III}^{-}$. In the bottom-right panel of Figure~\ref{fig:w_eff_sfree}, we observe that all trajectories asymptotically converge to this phantom solution, characterized by $w_{\rm eff} < -1$. This behavior is highlighted in the zoom-in panel (magenta curve). We also show a trajectory in which the universe initially evolves toward the (saddle) de Sitter point with $w_{\rm eff} = -1$, but eventually departs from it and settles into the phantom regime with $w_{\rm eff} < -1$.

Regarding initial conditions, we find that they primarily influence the early and intermediate stages of evolution, as expected, producing noticeable differences in the morphology of the curves before the Universe enters its accelerated phase. These effects are summarized in Fig. \ref{fig:w_eff_sfree}.

\begin{figure*}
\centering
\includegraphics[width=0.47\hsize,clip]{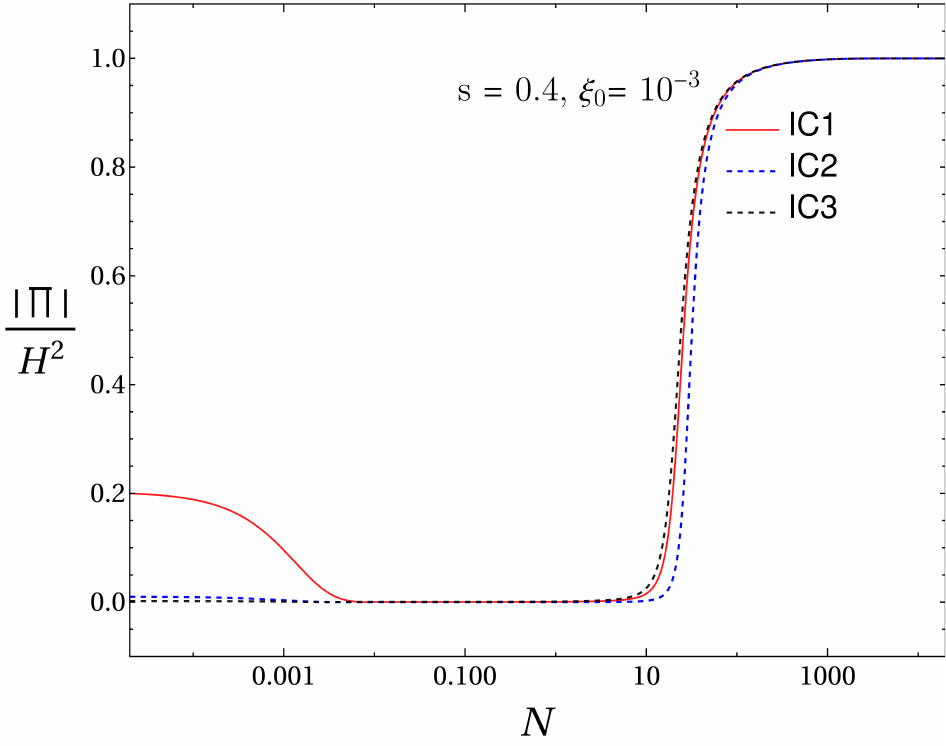}
\caption{Evolution of the ratio $|\Pi|/H^{2}$, which serves as an indicator of the relative contribution of viscous effects during cosmic expansion. We consider the case $s=0.4$ with $\xi_{0}=10^{-3}$ and explore three distinct initial conditions:  
IC1: $U(t_{\rm init}) = 0.5$, $V(t_{\rm init}) = 0.99$, $W(t_{\rm init}) = 0.2$;  IC2: $U(t_{\rm init}) = 0.7$, $V(t_{\rm init}) = 0.99$, $W(t_{\rm init}) = 0.01$; IC3: $U(t_{\rm init}) = 0.3$, $V(t_{\rm init}) = 0.01$, $W(t_{\rm init}) = 0.002$. The solid red curve (IC1) illustrates a scenario in which the effective dark matter pressure behaves as a stiff fluid due to the dominant contribution of the viscous pressure at early times. The two dashed curves (IC2 and IC3) correspond to cases in which viscosity becomes dynamically relevant only at late times, driving the accelerated expansion phase where $|\Pi|/p_{T} = 1$.}\label{fig:level_viscosity}
\end{figure*}

To quantify the impact of the viscous pressure throughout cosmic evolution, we plot the ratio $|\Pi|/H^{2}$, which provides a useful measure of the relevance of viscosity during the expansion history. This quantity serves as a robust estimator of the viscous contribution.\footnote{A more intuitive diagnostic would be the ratio $|\Pi|/p_{\text{T}}$, with $p_{T}$ being the total pressure. However, this expression becomes ill-defined whenever $p_{T} \to 0$, as occurs during matter domination, causing $|\Pi|/p_{T}$ to diverge.} 
In this sense, the ratio $|\Pi|/H^{2}$ becomes significant only when the dissipative fluid constitutes a significant fraction of the total energy budget, either at early times, when the fluid behaves effectively as a stiff component, or at late times, when it mimics dark energy. In these regimes, the fractional contribution of the viscous pressure can reach $\sim 20\%$ at early times and up to $100\%$ in the late-time asymptotic acceleration phase. This behavior is illustrated in Fig.~\ref{fig:level_viscosity}, where we have chosen the representative parameters $s = 0.4$, $v = 0.71$, $k = 0.3$, and $\xi_{0} = 10^{-3}$. As expected, the matter-dominated era remains essentially unaltered. However, increasing the bulk viscosity parameter $\xi_{0}$ leads to a departure from standard structure formation, potentially affecting the matter clustering amplitude $\sigma_{8}$. A detailed analysis of these effects is beyond the scope of this work but constitutes a promising direction for future investigation.

\subsection{The case $s=1/2$}

\begin{figure*}
\centering
\includegraphics[width=0.45\hsize,clip]{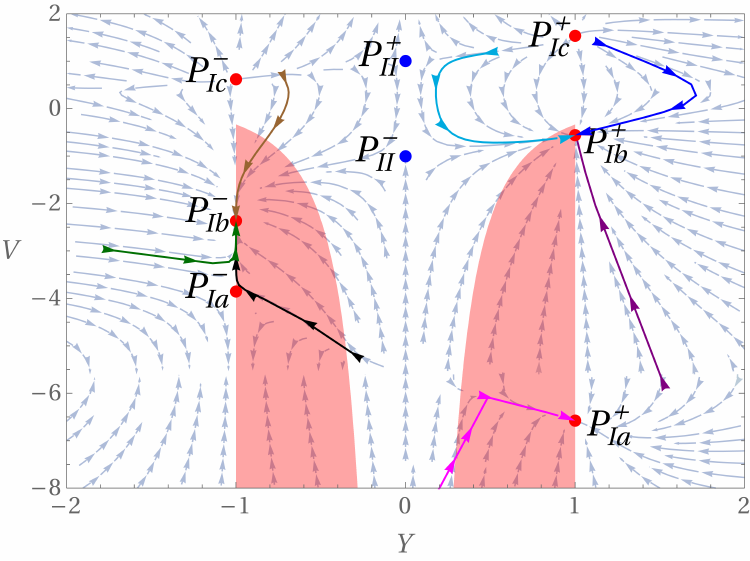}
\includegraphics[width=0.45\hsize,clip]{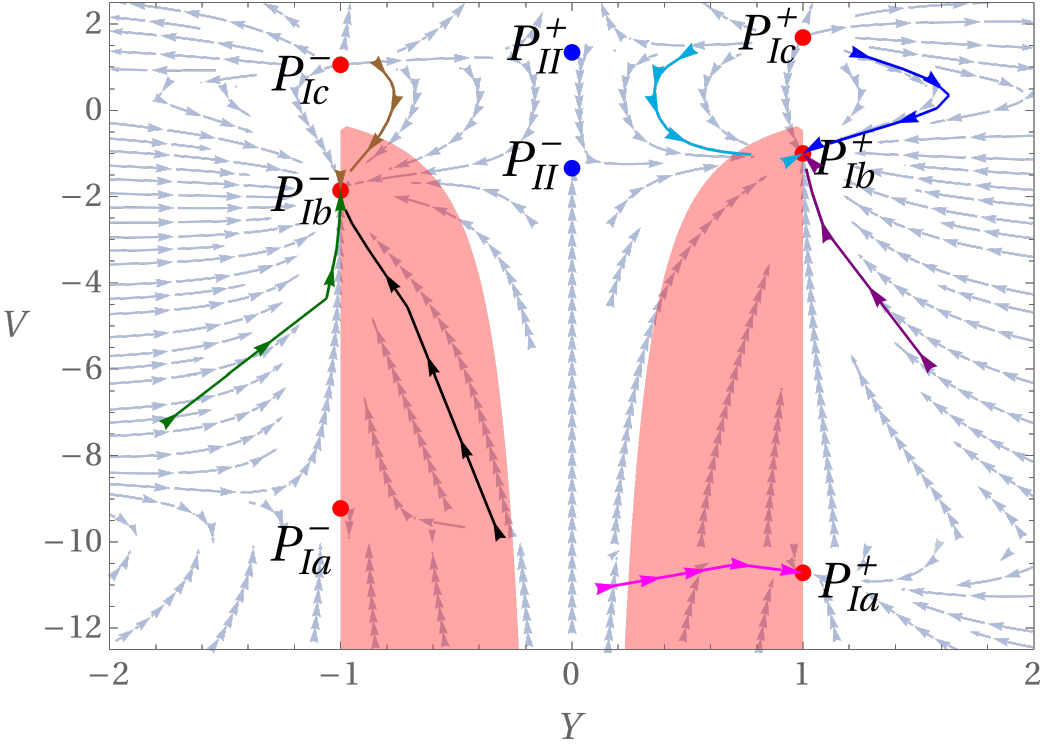}
\caption{Phase space portrait of the dynamical system for $s = 1/2$. Fixed points appear distributed nearly symmetrically with respect to $Y=0$, with similar dynamical character. For $Y = \pm 1$, three fixed points exist on each branch, labeled $P_{I}^{\pm}$. Among them, $P_{Ia}^{+}$, $P_{Ib}^{+}$, and $P_{Ib}^{-}$ all act as attractors within the physical region (light red). Two saddle points arises on each branch: $P_{Ic}^{+}$ and $P_{Ic}^{-}$. At $Y = 0$, two radiation-like points with $V_{II} = \pm \sqrt{2}v$ are present ($P_{II}^{\pm}$). Numerical trajectories illustrate the global behavior of the system. In the left panel, we adopt $v = 0.71$, $k = 0.3$, and $\xi_{0} = 0.9$, while in the right panel, we use $v = 0.95$, $k = 0.3$, and $\xi_{0} = 2.49$.} \label{fig:phase_space__shalf}
\end{figure*}

In Figure~\ref{fig:phase_space__shalf}, we display the phase space portrait of the dynamical system for $s = 1/2$. For illustrative purposes, we consider two representative limiting values of the non-adiabatic sound speed $v$, with $k$ held fixed. These values lie within the parameter space consistent with the existence of the fixed point and the acceleration conditions listed in Table~\ref{tab2:fixed_points_s_half}, as well as the stability properties summarized in Table~\ref{tab4::eigenvalues_s_half}. 
In the left top panel, we choose $v = 0.71$, $k = 0.3$, and treat $\xi_{0} = 0.9$ as a free parameter, representing a generic solution within the model. In the right top panel, we set $v = 0.95$, $k = 0.3$, and $\xi_{0} = 2.49$ where the right-side region ($Y>0$) coincides with the analytical solution in which $\xi_{0}$ is no longer arbitrary but is fixed by the values of the nonlinear parameters $k$ and $v$ (see Eq.~(\ref{bulk_coeff_plus})). More importantly, all the physical features of the generic solution are encapsulated in this analytical solution, as described below. The system is not fully symmetric with respect to the axis $Y = 0$; however, the same number of fixed points appears on both sides of this axis, each exhibiting the same dynamical character, with the exception of the points $P_{Ia}^{-}$ (a saddle) and $P_{Ia}^{+}$ (an attractor). 

For $Y = 1$, three fixed points emerge as solutions of the cubic\footnote{This is solved using a standard root-finding schema in \texttt{Mathematica}.} Eq.(\ref{polyn_V_s_onehalf}), corresponding to the positive branch, and are collectively denoted by $P_{I}^{+}$. One of them, labeled $P_{Ib}^{+}$, corresponds specifically to the analytical de Sitter-type solution $V_{I}^{*} = -\gamma$ ($\gamma=1$) when the bulk viscosity coefficient $\xi_{0}$ is determined by Eq.(\ref{bulk_coeff_plus}). In a more general case, it can behave like a quintessence field, as further discussed below. This point acts as an attractor and mimics a dark energy component driven by bulk viscosity, as does $P_{Ia}^{+}$. The latter, however, mimics a phantom-type dynamics, also acting as an attractor. In contrast, the point $P_{Ic}^{+}$ behaves as a saddle. Notably, the attractors lie within the physical region (highlighted in light red), where accelerated expansion is permitted. 

\begin{figure*}
\centering
\includegraphics[width=0.47\hsize,clip]{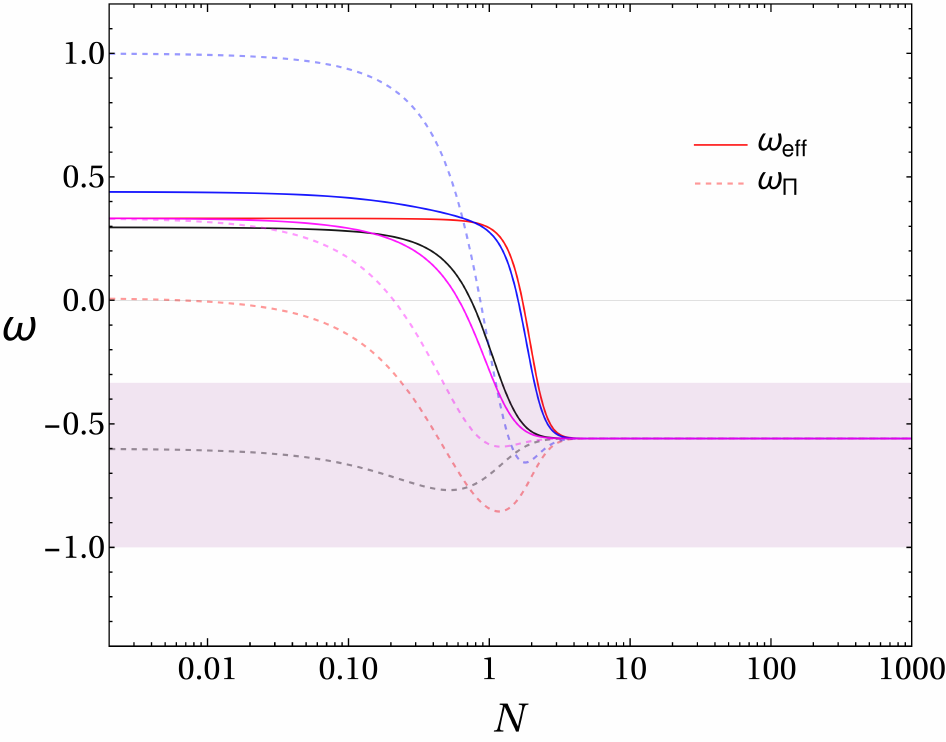}
\includegraphics[width=0.47\hsize,clip]{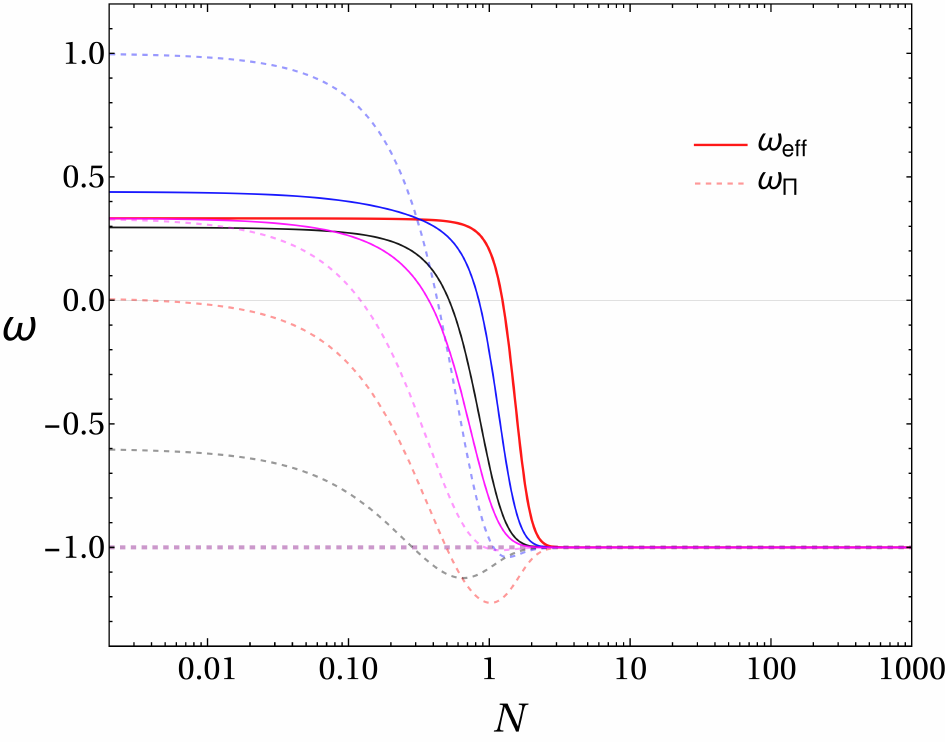}
\includegraphics[width=0.47\hsize,clip]{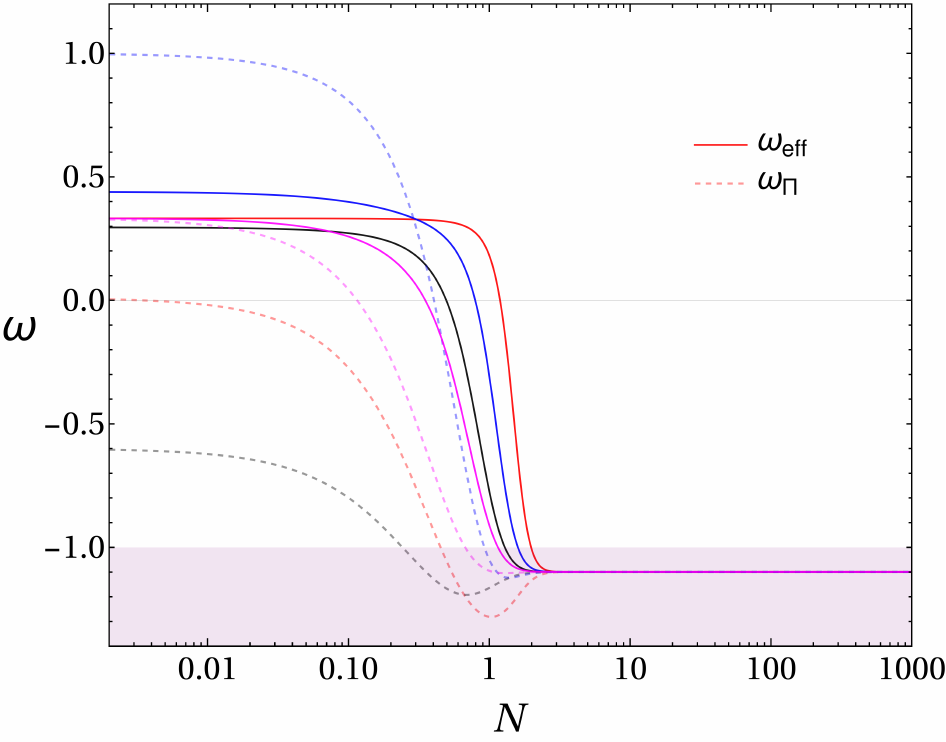}
\caption{Evolution of the effective EoS $w_{\rm eff}$ (solid curves) and the viscous DM EoS $w_{\Pi}$ (dashed curves) as functions of the number of e-folds $N$, for the case $s = 1/2$ and same initial conditions. \textit{Left top panel}: Numerical solutions for the generic case show that $w_{\Pi}$ evolves from an early dark energy–like phase ($w_{\Pi} \approx -0.6$) or stiff fluid behavior ($w_{\Pi} = 1$), eventually converging to $w_{\Pi} \approx -0.56$ at late times. The light purple region marks the range of quintessence-like behavior, $-1 \leq w_{\rm eff} < -1/3$, where all numerical solutions remain confined. For all cases we have adopted $v = 0.71$, $k = 0.3$, and $\xi_{0} = 0.9$. \textit{Right top panel:} Analytical solution in the constrained case, where $\xi_{0}$ is fixed by the non-linear parameters, and $w_{\Pi}$ approaches the de Sitter value $w_{\Pi} = -1$. Some trajectories briefly cross the phantom divide before settling into the late-time attractor. Here we use $v = 0.98$, $k = 0.3$, and $\xi_{0} = 2.4$. \textit{bottom panel}: This also depicts numerical solutions for the generic case. Contrary to the left top panel, we have used larger values of $v = 0.98$ and $\xi_{0} = 3.3$, while keeping $k = 0.3$ unchanged. These selected values drive both $w_{\rm eff}$ and $w_{\Pi}$ into an asymptotic phantom regime after crossing the phantom divide line. The corresponding region, where $w_{\rm eff} < -1$, is indicated by the light purple shading. In all plots, we have taken the following set of initial conditions: $Y(t_{\rm init}) = 0.05$, $V(t_{\rm init}) = 0.01$
for red curves; $Y(t_{\rm init}) = 0.2$, $V(t_{\rm init}) = -0.6$ for black curves; $Y(t_{\rm init}) = 0.4$, $V(t_{\rm init}) = 1$ for blue curves; and $Y(t_{\rm init}) = 0.5$, $V(t_{\rm init}) = 1/3$ for magenta curves.}\label{fig:weff_shalf}
\end{figure*}

A similar yet distinct structure arises for $Y = -1$, where the corresponding fixed points are collectively labeled as $P_{I}^{-}$. In contrast to the positive branch, these fixed points are solutions of the cubic equation~(\ref{polyn_V_s_Y_m_one_onehalf}). This case features a single attractor, denoted by $P_{Ib}^{-}$, along with two saddle points: $P_{Ia}^{-}$ and $P_{Ic}^{-}$. As in the positive branch, the attractor lies within the physical region of interest, supporting a phase of late-time accelerated expansion. We omit the numerical analysis of the specific analytical solution given by Eq.~(\ref{bulk_coeff_minus}), as it does not yield any new physical insights compared to the general case. For $Y = 0$, two radiation-like fixed points emerge with a non-trivial contribution of a viscous DM component, labeled as $P_{II}^{\pm}$. These behave as saddle points and are characterized by an effective viscous DM equation of state $w_{\Pi} = V_{II} = \pm \sqrt{2}v$, allowing for a wide range of early-time cosmological behaviors, as further illustrated in Figure~\ref{fig:weff_shalf}. 
Furthermore, the dynamical behavior near these fixed points is illustrated by several numerical trajectories initialized across a range of conditions that cover the relevant regions of the phase space. As expected, all trajectories ultimately converge to the attractor points.

There exists an intricate physical relation between the two panels depicted in Figure~\ref{fig:phase_space__shalf}. As the bulk viscosity coefficient $\xi_{0}$ increases, the system asymptotically approaches a de Sitter-like regime, characterized by the analytical solution $V_{I}^{*} = -\gamma$ ($\gamma=1$). This limiting behavior arises specifically when $\xi_{0}$ satisfies the condition given in Eq.~(\ref{bulk_coeff_plus}). In more general cases, however, the solution belongs to the broader family of quintessence cosmological models, defined by an effective EoS in the range $-1 \leq w_{\rm eff} < -1/3$. Within this family, the exact de Sitter solution with $w_{\rm eff} = -1$ appears as a special case.  

The general behavior of the model is captured by the effective EoS $w_{\rm eff}$ (solid curves) and the viscous DM EoS $w_{\Pi}$ (dashed light curves), as shown in Figure~\ref{fig:weff_shalf}. The left top panel displays numerical solutions of the system, showing that the viscous DM equation of state, $w_{\Pi}$, evolves through a broad spectrum of values during the early universe—from a dark energy-like component with $w_{\Pi} \approx -0.6$ to a stiff matter regime with $w_{\Pi} = 1$. Regardless of their early-time dynamics, all trajectories eventually converge toward a late-time value of $w_{\Pi} \approx -0.56$ for the chosen parameters. Notably, the analytical solution corresponds to the lower bound $w_{\Pi} = -1$, as illustrated in the right top panel. 

Finally, it is worth noting that some solutions exhibit a transient crossing of the phantom divide ($w_{\Pi} < -1$) before asymptotically approaching the de Sitter regime where $w_{\rm eff}$ settles (see right top panel). This behavior reflects the influence of bulk viscosity in temporarily driving $w_{\Pi}$ into the phantom domain, even when the long-term attractor is de Sitter–like. However, by increasing the model parameters to $v = 0.98$ and $\xi_{0} = 3.3$, it is possible for the trajectories to remain permanently in the phantom region after crossing the divide line. This globally impact the long-term attractor nature of $w_{\rm eff}$. In this case, the non-linear bulk-viscous effects dominate the late-time dynamics, preventing the system from relaxing back to the de Sitter state. This stable attractor phantom behavior is illustrated in the bottom panel of the same figure.

%


\subsection{Comparison with the Linear Israel--Stewart Theory}

\begin{figure*}
\centering
\includegraphics[width=0.47\hsize,clip]{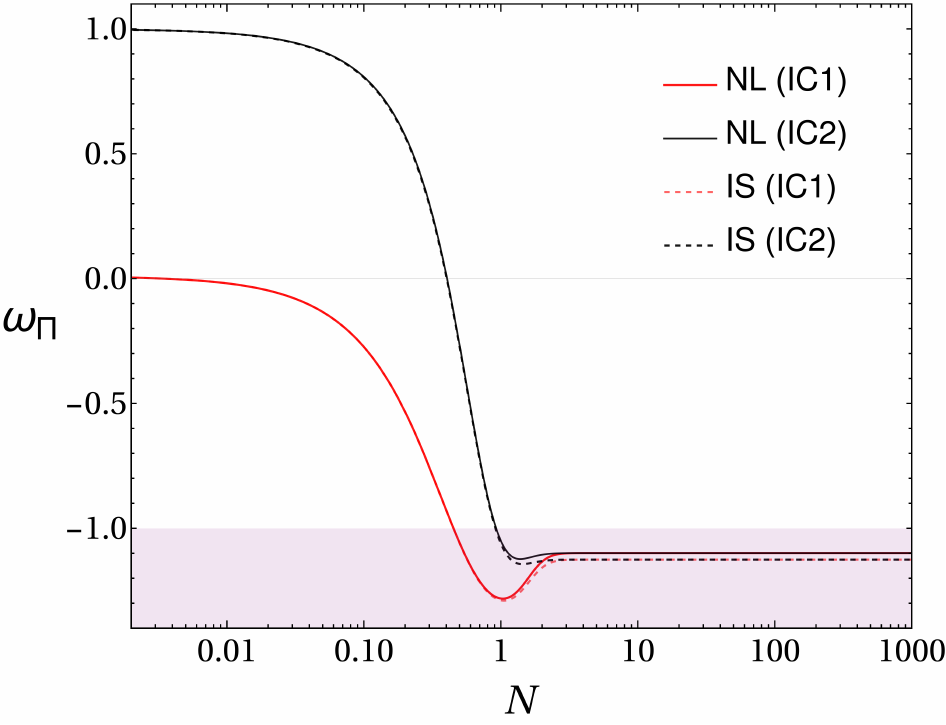}
\caption{Evolution of the viscous EoS $w_{\Pi}$ as a function of the number of e-folds $N$ for the representative case $s = 1/2$. Solid lines correspond to the non-linear model (NL) with $k = 0.3$, while dashed lines denote the linear Israel--Stewart (IS) theory obtained in the limit $k = 0$. In both cases we adopt the values $v = 0.98$ and $\xi_{0} = 3.3$. Moreover, for each framework we display two distinct sets of initial conditions: IC1, given by $Y(t_{\rm ini}) = 0.05$ and $V(t_{\rm ini}) = 0.01$, and IC2, given by $Y(t_{\rm ini}) = 0.4$ and $V(t_{\rm ini}) = 1$. The plot highlights the departure between the non-linear and IS descriptions at late times, where dissipative effects become dynamically relevant and non-linear corrections noticeably modify the evolution of $w_{\Pi}$.
}\label{fig:weff_compar}
\end{figure*}

There are important differences between the present non-linear viscous unified dark matter model and the standard linear Israel--Stewart (IS) theory that deserve to be emphasized. The motivation is primarily physical rather than merely quantitative: the non-linear formulation consistently accommodates far-from-equilibrium regimes characterized by $|\Pi| \gtrsim p_{m}$, where $p_{m}$ denotes the local equilibrium pressure of the fluid. Such regimes cannot be described within the linear IS theory, whose applicability is restricted to near-equilibrium configurations.
In a cosmological context---and particularly within unified dark matter scenarios—the far-from-equilibrium condition directly translates into the requirement for accelerated expansion,
\begin{equation}
\ddot{a} > 0 
\qquad \Longleftrightarrow \qquad 
|\Pi| > p_{T},
\end{equation}
where $p_{T}$ is the total equilibrium pressure. Thus, the non-linear extension provides a physically consistent framework for modeling viscous unified dark matter capable of driving cosmic acceleration, whereas the linear IS theory lies outside its domain of validity in this regime.\\
Let us momentarily set aside this limitation and focus on the quantitative differences between the two approaches. One might expect non-linear corrections to introduce only mild deviations, making both models nearly indistinguishable in regimes where the parameters $k$ and $v$ do not play a significant dynamical role. However, appreciable departures are expected whenever the cosmological solutions depend explicitly on these non-linear contributions. Recall that the IS theory is recovered in the limit $k = 0$, where all non-linear terms vanish in the autonomous system.\footnote{Although most IS critical points can be obtained as limiting cases of the non-linear solutions, it is technically safer to derive them directly from the linear autonomous system.}
These deviations are illustrated in Fig.~\ref{fig:weff_compar}, where we present the evolution of the viscous EoS $w_{\Pi}$ for the representative case $s = 1/2$, lying within the phantom regime. As anticipated, noticeable differences arise at late times. For example, the IS theory yields $w_{\Pi} = -1.12513$, while the non-linear theory gives $w_{\Pi} = -1.09896$. These correspond to fractional deviations of order $\sim 3\%$, which are not negligible in the context of high-precision cosmology.

%

\section{Conclusions}\label{sec:V}

Most existing bulk viscous unified DM models rely either on Eckart’s first-order theory—which suffers from well-known issues of causality and instability—or on the linear IS framework, which assumes small deviations from equilibrium. However, in the context of unified DM models, the IS theory is applied inconsistently, as it requires large dissipative pressures to drive cosmic acceleration, thereby violating its foundational assumptions. To address these limitations, we proposed a model based on a non-linear extension of the full causal IS theory, which is capable of handling large deviations from equilibrium arising from large bulk viscous stresses, i.e., $\Pi \gtrsim p$. This advancement highlights the necessity of adopting causal thermodynamics when modeling irreversible processes in the dark sector. It also opens the door to constructing realistic scenarios in which late-time acceleration emerges from within the DM fluid itself, without invoking an explicit dark energy component. Within this framework, we conducted a comprehensive analysis of unified DM models by combining dynamical system techniques and numerical analysis to characterize the full range of solutions, identify their stability properties, and determine the associated cosmological implications.

In our study we have adopted for a general framework for the analysis and presented particular analytical solutions that encapsulate the entire dynamics. We have also validate our analytical finding by numerically exploring the range of validity of the models parameters, which helped us to classify different cosmological scenarios and the underlying physical insights.

Our findings reveal that, for $s < 1/2$ and $v$ within the range $m < v^2 \leq 2 - \gamma$, with $m = Max \{ 1/2, k^2 \} $, a stable attractor fixed point ($P_{Ia}$) exists, as dictated by the stability analysis. This point corresponds to a purely viscous DM-dominated era with an EoS parameter $w_{\mathrm{eff}} = -1$, leading to an asymptotic bulk viscous pressure $\Pi_{\mathrm{as}} = -3 H^{2}{\infty} / (8 \pi G{N})$. This negative viscous pressure is, by itself, sufficient to drive the late-time accelerated expansion of the universe. Moreover, this expansion is asymptotically described by the de Sitter solution:
\begin{equation}
a(t) = a_0 \; \exp {\left( H_{\infty}\; t \right)}, \quad \text{with} \quad H_{\infty} = \sqrt{ \frac{3}{8 \pi G_N}} \; \left[ \frac{\xi_0}{v^2 \gamma} \;\left( v^2 - \frac{1}{2} \right) \left( 1 - \frac{k^2}{v^2} \right) \right]^{1/(1-2s)}.
\label{de_Sitter_solution} 
\end{equation}
In addition, and remarkably, this fixed point satisfies the condition of non-negative entropy production, as well as the positivity of the effective entropy (see, for instance, Eq.~(\ref{eff_entropy})).

The dissipative DM model described by the dynamical equations (\ref{evol_U})-(\ref{evol_W}) has another physically relevant fixed point for $s$ arbitrary:  $U_{II} = 0, V_{II} = \pm \sqrt{2} \; \gamma \; v, W_{II} = 0 $ (see Eq.~(\ref{FP_II})). This point corresponds to a pure radiation-dominated era with an effective equation of state parameter $w_{\rm eff} = 1/3$. Furthermore, since the dissipative bulk pressure vanishes ($\Pi = 0$), this fixed point describes a decelerated expansion with $ q = 1$. The existence of this second fixed point provides a potentially consistent framework for modeling the full cosmic history: beginning with a decelerated expansion during the radiation era, transitioning to the expected dark matter–dominated phase where bulk viscosity remains subdominant, and ultimately evolving into an accelerated expansion driven by the viscous DM pressure, without invoking an explicit dark energy component (see Fig.~\ref{fig:phase_space__sfree}).

Nevertheless, the late-time dynamics might face another distinctive stage for $s>1/2$: a phantom regime characterized by an effective equation of state $w_{\rm eff} = -\sqrt{2}v < -1$, associated with the fixed point $P_{III}^{-}$ (see discussion below Eq. (\ref{FP_III_phantom}), and bottom-right panel of Figure \ref{fig:w_eff_sfree}). Such a regime goes beyond the standard de Sitter expansion, naturally leading to the celebrated future singularities widely discussed in the literature.

Notably, between $P_{Ia}$ and $P_{III}^{-}$ lies the fixed point $P_{II}^{-}$, a basin-boundary saddle that separates two dynamical outcomes: 
direct convergence to the phantom attractor or a transient approach to the unstable de Sitter solution before ultimately reaching the phantom regime
(see Figure~\ref{fig:phase_space__sfree_phantom}). The presence of this basin-dividing saddle highlights the sensitivity of the dynamics to both initial conditions and parameter choices. Overall, this case reveals the rich phenomenology of non-linear bulk viscous unified DM models, where qualitatively distinct cosmic evolutions can emerge within a single theoretical framework.

Concerning the case $s = 1/2$, the dynamical system analysis reveals a qualitatively different structure compared to the generic $s \neq 1/2$ case. While the general scenario admits only a single late-time attractor—either de Sitter or phantom depending on the value of $s$, the $s=1/2$ case allows both solutions to exist simultaneously as attractors.  The de Sitter solution in this case appears as a particular limit of a more general quintessence-type dynamics (see Fig.~\ref{fig:phase_space__shalf}). This coexistence highlights a distinct stability structure, confirming that the $s=1/2$ case cannot be reduced to the generic scenario and must be treated separately.  

Overall, the nonlinear bulk viscosity scenario yields a rich and dynamically consistent cosmological picture in which viscous DM plays a central role throughout cosmic history. At early times, the viscous component can mimic a stiff fluid, while during intermediate epochs it behaves similarly to standard DM. At late times, bulk viscosity resurfaces as the dominant driver of accelerated expansion, giving rise to a spectrum of dark energy-like behaviors—ranging from quintessence and de Sitter phases to phantom-like regimes, as illustrated in Fig.~\ref{fig:weff_shalf}. These dynamics are governed by the nonlinear viscosity parameters $k$, $v$, and the coefficient $\xi_{0}$, which jointly shape the effective EoS and determine the asymptotic fate of the universe. 

Depending on the model parameters, the cosmological dynamics can exhibit a phantom crossing during an intermediate stage of the Universe's evolution. After crossing the phantom divide line, the solutions may either asymptotically approach a de Sitter state or evolve toward an eternal phantom regime, as illustrated numerically in Fig.~\ref{fig:weff_shalf}. These two possible asymptotic behaviors are separated by the basin-boundary saddle point $P_{II}^{\pm}$ (see Fig.~\ref{fig:phase_space__shalf}). In this sense, our results extend and generalize the findings of Ref.~\cite{Cataldo:2005qh}, providing a more consistent framework that goes beyond the restrictive assumptions of the standard Eckart and linear Israel–Stewart formalism.

It is important to emphasize that, within the framework of general relativity, phantom dark energy is typically considered unphysical due to violations of energy conditions and associated instabilities. In the present scenario, however, the phantom-like behavior is not an intrinsic property of the fluid but arises effectively from the non-equilibrium bulk viscous pressure. This interpretation preserves the physical consistency of the model while still allowing for rich late-time dynamics that go beyond the standard cosmological constant paradigm. 

A potential difficulty with unified DM models based on Eckart’s first-order theory of dissipation is that achieving a phase of accelerated cosmic expansion typically requires $\hat{\xi}_0 \sim \mathcal{O}(1)$, since $w_{\rm eff} = -\hat{\xi}_0$ (see Ref.~\cite{Palma:2024qrw}). This may conflict with the much smaller value ($\hat{\xi}_0 \sim 10^{-6}$) suggested by structure-formation considerations in Ref.~\cite{Anand:2017wsj} (also based on Eckart’s theory), although the latter does not strictly examine unified dark-matter scenarios. Such high values of $\hat{\xi}_0$ are also inconsistent with the present low cosmic DM density and with expectations based on the behavior of ordinary fluids, which exhibit small bulk viscosities under terrestrial environments. While the true microscopic nature of DM remains uncertain, it is highly unlikely that any viable DM model would support such large viscous coefficients. Moreover, successful structure formation requires the bulk viscosity to remain very small during the matter-dominated era \cite{Anand:2017wsj}, implying a counter-intuitive behavior of bulk viscosity diminishing as the energy density increases.  In contrast, our model, based on the full causal IS theory with a nonlinear extension, avoids this issue. Accordingly, adjusting $\xi_{0}$ primarily shifts the duration of the matter-dominated stage—higher values shorten it, while lower values prolong it—without altering the late-time attractor dynamics. This highlights the robustness of the accelerated expansion mechanism and the potential of the model in accommodating observational constraints on the matter-dominated era. We emphasize that the fixed points we identified in this framework persist, along with their stability properties, even for much smaller values of $\xi_0\sim 10^{-5}$. This demonstrates the robustness and physical plausibility of the model across a wide range of parameter values.

It is worth emphasizing that in unified DM models formulated within Eckart’s non-causal thermodynamics, the exponent $s$ must be restricted to $s>0$ in order to guarantee a physically viable cosmic evolution that includes a radiation-dominated stage \cite{Palma:2024qrw}. By contrast, in our non-linear causal framework, a radiation-dominated fixed point naturally exists for arbitrary values of $s$. Furthermore, while the sole attractor in the Eckart's theory corresponds to a de Sitter solution, the present model admits this solution only as a particular case within a broader family of quintessence-type attractors. 

Taken together, these results demonstrate that the limitations of Eckart-based models are not intrinsic to viscous unified DM scenarios themselves, but rather stem from the non-causal nature of Eckart’s approximation. The non-linear IS causal description, in contrast, provides a more general and physically robust framework, capable of encompassing radiation domination, quintessence-like acceleration, and even phantom-like asymptotics within a unified setting. 

Looking ahead, this framework offers compelling opportunities to address persistent observational tensions. In particular, the emergence of late-time acceleration from DM viscosity, rather than an independent dark energy field, presents an alternative route to easing the Hubble tension by modifying the expansion history in a physically motivated way. Moreover, the flexibility of the model to interpolate between different acceleration regimes aligns intriguingly with DESI’s recent evidence for evolving dark energy. A thorough, data-driven exploration, using constraints from SNe Ia, CMB, BAO and structure growth from current and future observational facilities, will be key to testing the viability of this scenario, constraining its parameters, and determining whether nonlinear viscous DM can serve as a unified description of both dark matter and dark energy phenomena.

\textbf{Data Availability Statement}: No Data associated in the manuscript.

\section*{Acknowledgments}
Financial support from the Chilean National Agency for Research and Development (ANID) through Fondecyt Grant No. 1250969 is gratefully acknowledged. We thank the anonymous referees for their valuable comments and suggestions.

\bibliography{biblio.bib}

\end{document}